\documentclass[a4paper,11pt]{article} 
\usepackage[utf8]{inputenc}
\usepackage[greek,english]{babel}
\languageattribute{greek}{ancient}
\usepackage{amsfonts,amsmath}
\usepackage{latexsym}
\usepackage{commath}
\usepackage{fontawesome}
\usepackage[makeroom]{cancel}
\usepackage{amssymb}
\usepackage{amsthm}
\usepackage{graphicx}
\usepackage{dsfont}
\usepackage{scalerel}
\usepackage[font=footnotesize]{caption}
\usepackage{bigints}
\usepackage[T2A,T1]{fontenc}
\usepackage{physics}
\usepackage[dvipsnames]{xcolor}
\usepackage{float}
\usepackage{authblk}
\usepackage{relsize,exscale}
\usepackage{mathtools}
\usepackage{xparse}
\usepackage{siunitx}
\usepackage{array}
\usepackage{etoolbox}
\usepackage{tensor}
\usepackage{nicefrac}

\usepackage{bm}
\usepackage[numbers,sort&compress]{natbib}
\usepackage{tocloft}  
\usepackage{spalign}
\usepackage{url}
\usepackage{microtype}
\usepackage{paralist}
\usepackage{wrapfig}
\usepackage{esint}
\usepackage{mathdots}
\usepackage{ffff}
\usepackage{titlesec}
\usepackage{hyperref}
\numberwithin{equation}{section}

\makeatletter
\newcommand*{\textoverline}[1]{$\overline{\hbox{#1}}\m@th$}
\makeatother
\DeclareFontEncoding{LS1}{}{}
\DeclareFontSubstitution{LS1}{stix}{m}{n}
\DeclareSymbolFont{stixletters}{LS1}{stix}{m}{it}
\DeclareMathAccent{\cev}{\mathord}{stixletters}{"91}
\DeclareMathAccent{\vec}{\mathord}{stixletters}{"92}
\DeclareMathAccent{\vecev}{\mathord}{stixletters}{"95}
\titleformat{\section}[block]{\large\bfseries\centering}{\thesection}{1em}{} 

\colorlet{myfilecolor}{violet}
\colorlet{myurlcolor}{Aquamarine}
\colorlet{mylinkcolor}{YellowOrange}
\definecolor{skobeloff}{rgb}{0.0, 0.48, 0.45}
\definecolor{hi}{RGB}{16, 9, 159}

\makeatletter
\newcommand{\megLeftrightarrow}[2][]{\ext@arrow 0099\Leftrightarrowfill@{#1}{#2}}
\makeatother

\DeclareFontFamily{U}{matha}{\hyphenchar\font45}
\DeclareFontShape{U}{matha}{m}{n}{
	<5> <6> <7> <8> <9> <10> gen * matha
	<10.95> matha10 <12> <14.4> <17.28> <20.74> <24.88> matha12
}{}
\DeclareSymbolFont{matha}{U}{matha}{m}{n}
\DeclareMathSymbol{\updownharpoons}{3}{matha}{"EA}
\DeclareMathSymbol{\downupharpoons}{3}{matha}{"EB}
\DeclareMathSymbol{\upharpoonleft}{3}{matha}{"E4}
\DeclareMathSymbol{\downharpoonright}{3}{matha}{"E7}
\DeclareFontFamily{U}{mathb}{\hyphenchar\font45}
\DeclareFontShape{U}{mathb}{m}{n}{
	<5> <6> <7> <8> <9> <10> gen * mathb
	<10.95> mathb10 <12> <14.4> <17.28> <20.74> <24.88> mathb12
}{}
\DeclareSymbolFont{mathb}{U}{mathb}{m}{n}
\DeclareMathSymbol{\upupharpoons}{3}{mathb}{"DA}
\DeclareMathSymbol{\downdownharpoons}{3}{mathb}{"DB}
\DeclareMathOperator\arctanh{arctanh}
\hypersetup{
	colorlinks=true,
	linkcolor=hi,
	filecolor=hi,      
	urlcolor=hi,
	citecolor=hi,
	linktoc=page,
	pdfnewwindow=true,
}
\makeatletter
\renewcommand*{\@fnsymbol}[1]{\ensuremath{\ifcase#1\or \natural \else\@arabic{\numexpr#1-\relax}\fi}}
\makeatother
\newcommand{\al}{\alpha}
\newcommand{\be}{\beta}
\newcommand{\g}{\gamma}
\newcommand{\de}{\delta}
\newcommand{\e}{\epsilon}

\newcommand{\la}{\lambda}
\newcommand{\m}{\mu}
\newcommand{\n}{\nu}

\newcommand{\s}{\sigma}

\newcommand{\ta}{\tau}

\newcommand{\om}{\omega}

\newcommand{\ten}{\tensor}

\DeclareSymbolFont{symbolstx}{OMS}{txsy}{m}{n}
\SetSymbolFont{symbolstx}{bold}{OMS}{txsy}{bx}{n}
\DeclareSymbolFontAlphabet{\mathcall}{symbolstx}

\DeclareMathOperator{\sgn}{sgn}

\def\xcpm#1#2#3{\mathbin{\ooalign{%
			\raise #2\hbox{\pdfliteral{1 0 0 rg}$#1+$\pdfliteral{0 g}}\cr
			\lower #3\hbox{\pdfliteral{0 0 1 rg}$#1-$\pdfliteral{0 g}}%
}}}
\title{\textbf{Massive gravity generalization of \(T\overline{T}\) deformations} }
\author{Evangelos Tsolakidis\textsuperscript{$\flat$,}\thanks{evangelos@hi.is \href{mailto:evangelos@hi.is}{\faEnvelopeO}}\vspace{0.2cm}}
\affil{\textsuperscript{$\flat$}\emph{Science Institute, University of Iceland,}\\ \emph{Dunhaga 3, 107 Reykjavík,} \\ \emph{Iceland}\vspace{-0.8cm}}
\date{}
\begin{document} 
	\maketitle
	\begin{abstract}
		\normalsize
		Motivated by the two-dimensional massive gravity description of \(T\overline{T}\) deformations, we propose a direct generalization in \(d\) dimensions. Our methodology indicates that all terms up to order \(d\) are present in the deformation. In two dimensions, \(T\overline{T}\) is enhanced by a linear and a constant term, and exhibits an interesting behaviour regarding the deformed spectrum and correlators. At certain limits, this deformation can reduce to \(T\overline{T}\) or \(T\overline{T}+\Lambda_{2}\) consistently. Using the massive gravity method, we obtain the classically deformed action of a sigma model of bosons and fermions interacting with an arbitrary potential, extending previous results. As a consequence, a proposal regarding the deformation of higher-derivative theories is made. Moreover, a standard dimensional reduction procedure is presented, with the resulting operator matching the form of prior findings under certain assumptions. In \(d\geq2\), we provide the exact structure of the quadratic terms in agreement with previous approaches, as well as the structure of the linear and constant terms. All higher order contributions are not easily evaluated, yet we derive the complete answer for all cases up to seven dimensions. Under certain conditions, these terms vanish, resulting in a quadratic operator. The trace-flow equation for this family of deformations is also derived. Finally,  we investigate the class of root-\(T\overline{T}\) operators in various dimensions within the scope of this formalism.     
	\end{abstract}
	\thispagestyle{empty}
	\newpage
\beforetochook\hrule
\tableofcontents
\afterTocSpace
\hrule
\afterTocRuleSpace
\flushbottom
\section{Introduction}
The systematic study of irrelevant deformations in two dimensions was sparked by Zamolodchikov's seminal discovery \cite{Zamolodchikov:2004ce}, and was established as an independent field of research over a decade later due to crucial advancements \cite{Smirnov:2016lqw,Cavaglia:2016oda}.  One of the main goals of this field is to provide a better understanding of the space that quantum field theories live in. This is closely connected to the flow equation   
\begin{gather}
	\pdv{S}{\la}=\mathcal{O}\,, \label{floweq}
\end{gather}
where \(S\) represents the action, \(\mathcal{O}\) the deforming operator and \(\la\) the parameter of the deformation. Usually, the deforming operator is taken to be the determinant of the stress-tensor or \(T\overline{T}\), due to its special quantum mechanical properties \cite{Zamolodchikov:2004ce}. 

Let us briefly motivate the equation above. The first thing we notice, is that \eqref{floweq} is equivalent to the standard way deformations are introduced under certain conditions. In more detail, if the deforming operator does not depend on \(\la\) it can be solved exactly yielding \(S_{0}+\la\mathcal{O}\) where \(S_{0}\) represents some seed quantum field theory. Clearly, when the deforming operator is a function of \(\la\) the solution is no longer trivial. The \(T\overline{T}\) operator falls into this category as it is constructed by the stress-tensor, which in turn is a function of the action \(S\) which implicitly depends on \(\la\). Since this flow equation holds in both cases, we understand that it is more fundamental, in a sense that it is the governing equation for deformations. Considering all of the above, it is interesting to note that the set of non-trivial solutions to this equation is generated by operators that depend on the action in some way, with one possible choice being the corresponding conserved currents of \(S\). 

 Since the discovery of the flow equation, various branches have sprouted in different research directions \cite{Bonelli:2018kik,Frolov:2019nrr,Coleman:2019dvf,Khoeini-Moghaddam:2020ymm,Caputa:2020lpa,Hou:2022csf,Bhattacharyya:2023gvg,Morone:2024ffm}. Notably, \(T\overline{T}\) can be understood as a random coordinate transformation \cite{Cardy:2018sdv}. Moreover, quite a few non-relativistic approaches exist \cite{Cardy:2018jho,Ceschin:2020jto,Esper:2021hfq,Hansen:2020hrs}. Besides \(T\overline{T}\),  the deformations constructed by currents other than the stress-tensor have also been studied, with the most common one being a \(U(1)\) symmetry current  \cite{Guica:2017lia,Frolov:2019xzi}, as well as deformations involving square roots \cite{Babaei-Aghbolagh:2022uij,Ferko:2022cix,Babaei-Aghbolagh:2022leo,Hadasz:2024pew}.  From a holographic point of view, a \(T\overline{T}\)-deformed conformal field theory on the boundary was found to be dual to \(\text{AdS}_{3}\) at finite radial cut-off \cite{McGough:2016lol,Guica:2019nzm} with various generalizations developed in \cite{Bzowski:2018pcy,Gorbenko:2018oov,Kraus:2018xrn,Poddar:2023ljf}. Interestingly, the radial cut-off was found to be proportional to the inverse square of the deformation parameter. Another intriguing reformulation of \(T\overline{T}\) was discovered in \cite{Dubovsky:2017cnj,Dubovsky:2018bmo}, where the deformed theory was found to be equivalent to flat space Jackiw-Teitelboim gravity coupled to the undeformed one. A curved background generalization exists \cite{Tolley:2019nmm, Mazenc:2019cfg}, where the undeformed action couples to massive gravity instead. Furthermore,  \(T\overline{T}\) deformations of  correlators have been investigated \cite{Kraus:2018xrn,Cardy:2019qao,Cui:2023jrb,Aharony:2023dod,Barel:2024dgv}. In higher dimensions, the structure of these stress-tensor deformations is an open problem, yet there are certain proposals of quadratic order   \cite{Taylor:2018xcy,Hartman:2018tkw,Conti:2022egv,Ondo:2022zgf,Ebert:2024fpc}. Finally, a deformation in one dimension can be derived via dimensional reduction of \(T\overline{T}\) \cite{Gross:2019ach}.
 
 In this work, we will mostly focus on the massive gravity\footnote{For further details on massive gravity we refer the interested reader to \cite{Hinterbichler:2012cn, deRham:2014zqa}.} approach \cite{Tolley:2019nmm, Mazenc:2019cfg}, extending its formalism in \(d\) dimensions. We were motivated to adopt this method for the study of deformations, due to the fact that the seed theory does not need to be specified. We begin by developing the general methodology in \(d\) dimensions. By introducing a background and an auxiliary vielbein, we assume that the solution of \eqref{floweq} can be written as \(S_{\text{G}}+S_{0}\), where \(S_{\text{G}}\) depends on both vielbeins and \(\la\) whereas \(S_{0}\) only on the auxiliary vielbein and other fields. Once a deformation (which is a function of the stress-tensor) is specified, the flow equation determines the exact expression for \(S_{\text{G}}\). When \(S_{\text{G}}+S_{0}\) is evaluated on-shell for the auxiliary vielbein, the resulting expression is the deformed action. In \(d=2\), we generalize the massive gravity method \cite{Tolley:2019nmm,Mazenc:2019cfg} by considering an operator quadratic in the stress-tensor, that reduces correctly to \(T\overline{T}\) as well as  \(T\overline{T}+\Lambda_{2}\) in certain limits. The spectrum of this operator can be obtained, and an identification with a string theory spectrum is evident. Moreover, the structure of the linear term allows for the calculation of certain deformed correlators. A dimensional reduction of the deformation yields a one-dimensional operator, which agrees with previous proposals \cite{Gross:2019ach} under certain assumptions. As another application of this methodology, we provide an expression for the  classically deformed sigma model of bosons and fermions interacting with an arbitrary potential, extending previous results. We find that the resulting expression allows for the deformation of certain higher-derivative theories. Finally, root-\(T\overline{T}\) is discussed from the massive gravity perspective. In a complete analogy to the two-dimensional case, we extend this approach in \(d\) dimensions. Specifying the deforming operator to be an order \(d\) polynomial in the stress-tensor, yields a quadratic structure that precisely matches previous treatments \cite{Taylor:2018xcy,Hartman:2018tkw}. All higher order terms were too cumbersome to determine for general \(d\), yet we do provide the complete expression of the deforming operator for all cases up to seven dimensions in appendix \ref{bees}. Lastly, we investigate the properties of root-\(T\overline{T}\) through the lens of massive gravity.

This paper is structured in the following way. In section \ref{sec2}, we develop the massive gravity formalism in \(d\) dimensions. During section \ref{main2d}, we apply our methodology in \(d=2\) recovering and partially extending many known results. The deforming operators in three, four and \(d\) dimensions are derived in section \ref{sechigh}, followed by a small discussion on root-\(T\overline{T}\). Finally, in section \ref{disc} we provide a detailed summary of our work, along with some arguments on the potential constraints imposed on the stress-tensor, that result in an operator of quadratic order in \(d\) dimensions. We conclude with some discussions and a possible generalization of our method.\\

\noindent\textbf{Note added:} During the preparation of this manuscript, \cite{Babaei-Aghbolagh:2024hti} appeared on the arXiv, where partial overlap is noted with section \ref{sec2}, as well as subsections \ref{secroot}, \ref{sec3d} and \ref{sec4d} of our paper.
\section{Methodology in \(d\) dimensions} \label{sec2}
Let us now develop the algorithm that will be extensively used in the upcoming sections. This approach shares some common ground with the ansatz method, which is developed in appendix \ref{ansatz} for comparison reasons. Following the title of this section, we will temporarily work in \(d\) dimensions. As in \cite{Tolley:2019nmm, Mazenc:2019cfg} we will use two dimensionless vielbeins, the auxiliary \(\tensor{e}{_{\m}^{a}}\) and its corresponding metric \(\tensor{g}{_{\m}_{\n}}=\tensor{e}{_{\m}^{a}}\tensor{e}{_{\n}^{b}}\tensor{\eta}{^{a}^{b}}\), as well as the background vielbein \(\tensor{f}{_{\m}^{a}}\) together with \(\tensor{\g}{_{\m}_{\n}}=\tensor{f}{_{\m}^{a}}\tensor{f}{_{\n}^{b}}\tensor{\eta}{^{a}^{b}}\). It will also be convenient to define the quantities \(Y_{i}\) and \(y_{i}\) as 
\begin{gather}
	\tensor{\qty(Y_{i})}{^{\m}_{\n}}\coloneqq\tensor{f}{^{\m}_{a_{1}}}\tensor{e}{_{\la_{1}}^{a_{1}}}\tensor{f}{^{\la_{1}}_{a_{2}}}\tensor{e}{_{\la_{2}}^{a_{2}}}\cdots\tensor{f}{^{\la_{i-1}}_{a_{i}}}\tensor{e}{_{\n}^{a_{i}}}\,, \qquad y_{i}\coloneqq\tensor{\qty(Y_{i})}{^{\m}_{\m}}=\tr(Y_{i})\,, \label{genop}
\end{gather}
that obey equations \eqref{eq11}. Now consider an operator \(\mathcal{O}\) that satisfies the standard flow equation \eqref{floweq}, and further assume that it is a function of the deformation parameter \(\la\) and potentially other couplings \(\mathfrak{g}_{i}\), that it functionally depends on the stress-tensor and of course the background \(\tensor{f}{_{\m}^{a}}\). In other words 
\begin{gather}
	\mathcal{O}=\int\dd[d]{x}f\hspace{0.05cm}F(\la,\mathfrak{g}_{i},\tensor{T}{^{\m}_{a}})\,, \qquad \tensor{T}{^{\m}_{a}}\coloneqq\dfrac{1}{f}\dfrac{\de S}{\de \tensor{f}{_{\m}^{a}}}\,,
\end{gather}
where \(f\coloneqq\det\tensor{f}{_{\m}^{a}}\) and \(F\) is a function that one specifies according to the desired deformation. Then, via defining \(W_{i}\) and \(w_{i}\) in a similar manner
\begin{gather}
	\tensor{\qty(W_{i})}{^{\m}_{\n}}\coloneqq\tensor{T}{^{\m}_{a_{1}}}\tensor{f}{_{\la_{1}}^{a_{1}}}\tensor{T}{^{\la_{1}}_{a_{2}}}\tensor{f}{_{\la_{2}}^{a_{2}}}\cdots\tensor{T}{^{\la_{i-1}}_{a_{i}}}\tensor{f}{_{\n}^{a_{i}}}\,, \qquad w_{i}\coloneqq\tensor{\qty(W_{i})}{^{\m}_{\m}}=\tr(W_{i})\,,
\end{gather} 
and using the Cayley-Hamilton theorem \eqref{eqa7}, \(F\) is restricted to \(F(\la,\mathfrak{g}_{i},w_{1},\dots,w_{d})\). Consider now the following expression
\begin{gather}
	S[\tensor{e}{_{\m}^{a}},\tensor{f}{_{\m}^{a}},\la,\mathfrak{g}_{i},\Phi]=\int\dd[d]{x}f\hspace{0.05cm}G(\la,\mathfrak{g}_{i},y_{1},\dots,y_{d})+S_{0}[\tensor{e}{_{\m}^{a}},\mathfrak{g}_{i},\Phi]\,, \label{sol1}
\end{gather}
which is the solution to the flow equation when evaluated on-shell for \(\tensor{e}{_{\m}^{a}}\), at least classically. We will usually refer to the first term of the action above as the massive gravity action or \(S_{\text{G}}\). The function \(G\) is any solution of the flow equation which takes the form of a first order non-linear \((d+1)\)-dimensional partial differential equation and \(S_{0}\) is the seed action,\footnote{This action is arbitrary, it may contain curvature terms. The variable \(\Phi\) represents matter.} which is usually (but not always) the part that survives the \(\la\rightarrow 0 \) limit. Also notice that \(G\) is restricted in the same way as \(F\) for the same reasons. Finally, it is evident that the nature of the seed action comes into play only when the equations of motion for the auxiliary vielbein are considered.

In order to better understand the origin of this solution consider a fixed \(F\). Now, one can follow with the variation of \eqref{sol1} with respect to \(\tensor{f}{_{\m}^{a}}\) from which it is obvious that only the function \(G\) will survive. It is important to take these variations before evaluating everything on-shell for the auxiliary vielbein.  A simple chain rule and we reach
\begin{gather}
	W_{1}=G\mathds{1}_{d}-\sum_{j=1}^{d}j\pdv{G}{y_{j}}Y_{j}\,, \label{defomten}
\end{gather}
where \(\mathds{1}_{d}\) is the \(d\)-dimensional identity matrix, and
\begin{gather}
	w_{i}=dG^{i}+\sum_{k=1}^{i}(-1)^{k}G^{i-k}\binom{i}{k}\sum_{j_{1},\dots,j_{k}=1}^{d}\qty[\prod_{n=1}^{k}j_{n}\pdv{G}{y_{j_{n}}}]y_{j_{1}+\cdots+j_{k}}\,, \label{wa}
\end{gather}
where \(y_{j_{1}+\cdots+j_{k}}\) can be written as a function of \((y_{1},\dots,y_{d})\), again from \eqref{eqa7}. Now assuming that the on-shell condition for the auxiliary vielbein holds, the flow equation \eqref{floweq} simply reads
\begin{gather}
	\partial_{\la}G(\la,\mathfrak{g}_{i},y_{1},\dots,y_{d})=F(\la,\mathfrak{g}_{i},w_{1},\dots,w_{d})\,, \label{pde}
\end{gather}
i.e. the first order non-linear \((d+1)\)-dimensional partial differential equation for \(G\). It is important to stress that a priori, one does not need to find the complete solution to \eqref{pde}, any function that satisfies it is a potential candidate. The major restriction to the set of possible solutions comes from the \(\la\rightarrow 0\) limit as well as dimensional analysis. The current understanding of the limit is simple, one recovers the seed action \(S_{0}\) on the \(\tensor{f}{_{\m}^{a}}\) background. In mathematical terms this translates to the following initial condition  
\begin{gather}
\tensor{e}{_{\m}^{a}}=\tensor{f}{_{\m}^{a}}\,, \qquad	G(0,\mathfrak{g}_{i},d,\dots,d)=0\,, \label{inco}
\end{gather}
where the first equation should drop out naturally from the equations of motion for the auxiliary vielbein in that limit. During the next section we will see that one may relax the initial condition above in a specific way. Moreover, since the vielbeins are assumed to be dimensionless and \(G\) has mass units \(M^{d}\), the explicit dependence on \(\la\) and \(\mathfrak{g}_{i}\) is fixed by dimensional analysis.

Furthermore, the equations of motion of the auxiliary vielbein for \eqref{sol1} can be evaluated yielding
\begin{gather}
	\tensor{\bar{\mathcall{S}}}{^{\mu}_{a}}\tensor{\bar{e}}{_{\nu}^{a}}=-\dfrac{1}{\det \bar{Y}_{1}}\sum_{j=1}^{d}j\pdv{\bar{G}}{\bar{y}_{j}}\tensor{\qty(\bar{Y}_{j})}{^{\m}_{\n}}\,, \qquad  \label{eqmot} 	\tensor{\mathcall{S}}{^{\mu}_{a}}\coloneqq\dfrac{1}{e}\dfrac{\de S_{0}}{\de \tensor{e}{_{\m}^{a}}}\,,
\end{gather}
where \(\tensor{\bar{e}}{_{\mu}^{a}}\) is the solution, \(e\coloneqq\det\tensor{e}{_{\m}^{a}}\),  and the bar denotes all on-shell affected quantities. Once more, we define \(Z_{i}\) as
\begin{gather}
		\tensor{\qty(Z_{i})}{^{\m}_{\n}}\coloneqq\tensor{\mathcall{S}}{^{\m}_{a_{1}}}\tensor{e}{_{\la_{1}}^{a_{1}}}\tensor{\mathcall{S}}{^{\la_{1}}_{a_{2}}}\tensor{e}{_{\la_{2}}^{a_{2}}}\cdots\tensor{\mathcall{S}}{^{\la_{i-1}}_{a_{i}}}\tensor{e}{_{\n}^{a_{i}}}\,, \qquad z_{i}\coloneqq\tensor{\qty(Z_{i})}{^{\m}_{\m}}=\tr(Z_{i})\,,
\end{gather}
and together with \eqref{defomten} we reach
\begin{gather}
		f\bar{W}_{1}=f\bar{G}\mathds{1}_{d}+\bar{e}\bar{Z}_{1}\,.
\end{gather}
The equation above essentially relates the deformed stress-tensor density with the undeformed one when the Einstein-Hilbert term is absent. As expected, when \(\la\rightarrow 0\) and \eqref{inco} holds the stress-tensors match trivially.

To summarise, taking into consideration all of the above, we naturally reach the following, relatively simple algorithm: First, one begins by fixing \(F\). Then the evaluation of \(w_{i}\) follows, using \eqref{sol1} and \eqref{wa}. This results in a first order non-linear \((d+1)\)-dimensional partial differential equation for \(G\) i.e. \eqref{pde}, which in principle can be solved using e.g. the method of characteristics or an ansatz. In order to obtain the full solution, the initial condition \eqref{inco} is used followed by the elimination of \(\tensor{e}{_{\m}^{a}}\) via \eqref{eqmot}. This step is classical, one may choose to perform the path integral over the auxiliary vielbein. Finally, one can apply this method iteratively. In more detail, for each iteration the background vielbein becomes the new auxiliary vielbein and the action becomes the new seed action. A more schematic explanation is presented in figure \ref{fig1} bellow.
\begin{figure}[H]
	\centering
\begin{gather*}
	\begin{gathered}
	S^{(0)}_{0}[\tensor{{e_{0}}}{_{\m}^{a}},\mathfrak{g}_{i},\Phi]\overunderset{\mathcal{O}_{1,\la_{1}}}{\tensor{{e_{1}}}{_{\m}^{a}}}{\xrightarrow{\hspace*{1cm}}}\cdots\overunderset{\mathcal{O}_{n-1,\la_{n-1}}}{\tensor{{e_{n-1}}}{_{\m}^{a}}}{\xrightarrow{\hspace*{1.5cm}}} S^{(n-1)}_{0}[\tensor{{e_{0}}}{_{\m}^{a}},\dots,\tensor{{e_{n-1}}}{_{\m}^{a}},\la_{1},\dots,\la_{n-1},\mathfrak{g}_{i},\Phi]\overunderset{\mathcal{O}_{n,\la_{n}}}{\tensor{f}{_{\m}^{a}}}{\xrightarrow{\hspace*{1cm}}}\\\overunderset{\mathcal{O}_{n,\la_{n}}}{\tensor{f}{_{\m}^{a}}}{\xrightarrow{\hspace*{1cm}}} S[\tensor{{e_{0}}}{_{\m}^{a}},\dots,\tensor{{e_{n-1}}}{_{\m}^{a}},\tensor{f}{_{\m}^{a}},\la_{1},\dots,\la_{n},\mathfrak{g}_{i},\Phi]
\end{gathered}
\end{gather*}
\caption{Iteration of \(n\), not necessarily identical deformations using the massive gravity method. At each step one is free to integrate out all or part of the auxiliary vielbeins at will.}
\label{fig1}
\end{figure}  
\noindent We will mostly work out the first four steps in various dimensions, leaving the rest for future endeavours. 
\section{Two dimensions} \label{main2d}
Shifting our attention to two dimensions, we will begin by deriving the deforming operator and check that at certain limits, one gets the correct massive gravity action for the well-studied \(T\overline{T}\) operator \cite{Tolley:2019nmm, Mazenc:2019cfg}, as well as the \(T\overline{T}+\Lambda_{2}\) operator \cite{Torroba:2022jrk}. Then, we will briefly study its spectrum and observe that it can be identified with a more general string theory spectrum. A small discussion about the deformed correlators of undeformed operators will follow, as well as a dimensional reduction to \(d=1\). We will then derive the classically deformed theories for a sigma model of bosons and fermions interacting with an arbitrary potential in one and two dimensions and propose a way to extend the deformation to higher derivative theories. Finally, we will close this section with a study of root-\(T\overline{T}\) \cite{Ferko:2022cix} via the massive gravity method.    
\subsection{Quadratic deformations}
 We start from the generic solution \eqref{sol1}, which now reads
\begin{gather}
	S[\tensor{e}{_{\m}^{a}},\tensor{f}{_{\m}^{a}},\la,\mathfrak{g}_{i},\Phi]=\int\dd[2]{x}f\hspace{0.05cm}G(\la,\mathfrak{g}_{i},y_{1},y_{2})+S_{0}[\tensor{e}{_{\m}^{a}},\mathfrak{g}_{i},\Phi]\,. \label{osol2}
\end{gather}
Following the first step of the algorithm, we specify the deformation (which is irrelevant in this case) to be of quadratic order i.e.
\begin{gather}
	\mathcal{O}=\int\dd[2]{x}f\hspace{0.05cm}\qty{b_{0}\tensor{T}{^\mu_\nu}\tensor{T}{^\nu_\mu}+b_{1}(\tensor{T}{^\mu_\mu})^{2}+\frac{b_{2}}{\la}\tensor{T}{^\mu_\mu}+\frac{b_{3}}{\la^{2}}}\,, \label{2ddef}
\end{gather}
where the dependence\footnote{It is important to note that this choice is unique only when no other dimensionful couplings are present. It is understood that there do exist infinite combinations of more than one couplings that carry the correct units.} on \(\la\) is such that \(b_{0}, b_{1}, b_{2}, b_{3}\) are dimensionless arbitrary constants. Then, \eqref{pde} and \eqref{2ddef} yield the corresponding flow equation, which we do not explicitly state since it is quite lengthy and does not contribute to our understanding. Nevertheless, we can still solve it through a quadratic ansatz for \(G\)
\begin{gather}
	G(\la,\mathfrak{g}_{i},y_{1},y_{2})=c_{0}(\la,\mathfrak{g}_{i})+c_{1}(\la,\mathfrak{g}_{i})y_{1}+c_{2}(\la,\mathfrak{g}_{i})y_{1}^{2}+c_{3}(\la,\mathfrak{g}_{i})y_{2}\,, \label{gsol}
\end{gather}
which yields the following system of equations
\begin{gather}
	\begin{gathered}
		b_{0}(c_{3}+c_{2})=0\,, \quad \partial_{\la}c_{3}=b_{0}c_{1}^{2}\,, \quad \partial_{\la}c_{2}=b_{1}c_{1}^{2}\,, \quad \partial_{\la}c_{1}=\dfrac{c_{1} (2\la c_{0}(b_{0}+2 b_{1})+b_2)}{\la}\,,\\ \partial_{\la}c_{0}=\dfrac{2\la c_{0}(\la c_{0}(b_{0}+2 b_{1})+b_{2})+b_{3}}{\la^{2}}\,,
	\end{gathered} \label{2dpro}
\end{gather}
where the dependences have been dropped for notational economy. The system is overdetermined but the last four equations can be easily solved i.e.
\begin{gather}
\begin{gathered}
\begin{aligned}
	c_{0}(\la,\mathfrak{g}_{i})&=\frac{b\qty(\frac{2\mathfrak{c_{0}}(\mathfrak{g}_{i})}{\la^{b}+\mathfrak{c_{0}}(\mathfrak{g}_{i})}-1)-2b_{2}-1}{4\la\qty(b_{0}+2b_{1})}\,, 
	&c_{1}(\la,\mathfrak{g}_{i})&=\frac{\sqrt{\la}^{b-1}\mathfrak{c_{1}}(\mathfrak{g}_{i})}{\la^{b}+\mathfrak{c_{0}}(\mathfrak{g}_{i})}\,,\\
	c_{2}(\la,\mathfrak{g}_{i})&=\mathfrak{c_{2}}(\mathfrak{g}_{i})-\frac{b_{1}\mathfrak{c_{1}}(\mathfrak{g}_{i})^{2}}{b\qty(\la^{b}+\mathfrak{c_{0}}(\mathfrak{g}_{i}))}\,,  &c_{3}(\la,\mathfrak{g}_{i})&=\mathfrak{c_{3}}(\mathfrak{g}_{i})-\frac{b_{0}\mathfrak{c_{1}}(\mathfrak{g}_{i})^{2}}{b\qty(\la^{b}+\mathfrak{c_{0}}(\mathfrak{g}_{i}))}\,,
\end{aligned}\\[0.3cm]
b\coloneqq\sqrt{(1+2b_{2})^{2}-8b_{3}(b_{0}+2b_{1})}\,,
\end{gathered} \label{2dsol}
\end{gather}
where \(\mathfrak{c_{0}}, \mathfrak{c_{1}}, \mathfrak{c_{2}}, \mathfrak{c_{3}}\) are integration constants that depend exclusively on other couplings of the undeformed theory. Now, revisiting  the first equation of \eqref{2dpro} we distinguish two independent cases: \(b_{0}=0\) or \(c_{3}=-c_{2}\). In the first case, the solution is given by \eqref{2dsol} in that limit, but in the second, we get that
\begin{gather}
	b_{1}=-b_{0}\,, \qquad \mathfrak{c_{3}}=-\mathfrak{c_{2}}\,. \label{con1}
\end{gather}
With this choice and looking back to \eqref{2ddef}, we notice that the first two terms of the deformation are analogous to the \(T\overline{T}\) operator. During the next several subsections, we will focus exclusively on this case. 
 
Let us now analyse the result above further. As previously mentioned in section \ref{sec2}, the function \(G\) has mass units \(M^{2}\) in two dimensions, which in turn translates to the units of the coefficients \(c_{i}\) since the zweibeins are unitless. Furthermore for \eqref{2ddef}, \(\la\) has mass units \(M^{-2}\) therefore from \eqref{2dsol} the units of \(\mathfrak{c_{i}}\) are uniquely determined or
\begin{gather}
	\qty[\mathfrak{c_{0}}(\mathfrak{g}_{i})]=M^{-2b}\,, \quad \qty[\mathfrak{c_{1}}(\mathfrak{g}_{i})]=M^{1-b}\,,  \quad  \qty[\mathfrak{c_{2}}(\mathfrak{g}_{i})]=M^{2}\,,  \quad  \qty[\mathfrak{c_{3}}(\mathfrak{g}_{i})]=M^{2}\,, \label{diman}
\end{gather}
which means that these constants are actually constrained functions of the other couplings of the seed theory. 
\subsubsection{Retrieving \(T\overline{T}\)}
Considering now \eqref{con1}, the standard massive gravity description of \(T\overline{T}\) naturally drops out when the coefficients \(\mathfrak{c_{i}}\) do not depend on other couplings i.e. they are dimensionless. The combination of this fact and \eqref{diman} uniquely leads to \(\mathfrak{c_{0}}=0=\mathfrak{c_{2}}\) and \(b=1\), which in turn translates to the following constraint
\begin{gather}
	b_{3}=-\frac{b_{2}(1+b_{2})}{2b_{0}}\,. \label{con2}
\end{gather}
The last thing that needs to be fixed is the constant \(\mathfrak{c_{1}}\), which is determined simply via \eqref{inco} in accordance with \eqref{eqmot}. We find that \(b_{2}=0\) and \(\mathfrak{c_{1}}=-1/(2b_{0})\). In order to generalize this result let us relax \eqref{inco} by allowing \(b_{2}\) to be arbitrary. Then, the massive gravity action reads 
\begin{gather}
	S_{\text{G}}=\int\dd[2]{x}\dfrac{f}{\la'}\qty{1+b_{2}-y_{1}+\det Y_{1}}=\int\dd[2]{x}\dfrac{f}{\la'}\qty{1+b_{2}-y_{1}}+\int\dd[2]{x}\dfrac{e}{\la'}\,, \label{soll2d}
\end{gather}
where \(\la'\coloneqq2\la b_{0}\), and similarly \eqref{2ddef} becomes
\begin{gather}
		\mathcal{O}=\int\dd[2]{x}f\hspace{0.05cm}2b_{0}\qty{-\det\tensor{T}{^\mu_\nu}+\frac{b_{2}}{\la'}\tensor{T}{^\mu_\mu}-\frac{b_{2}(1+b_{2})}{\la'^{2}}}\,. \label{opp2d}
\end{gather}
Now the \(\la\rightarrow 0\) limit is not well-defined, in fact it diverges. We will attempt to provide an explanation of this behaviour in the upcoming sections. Finally, turning off \(b_{2}\) and making use of \eqref{levci} we retrieve
\begin{gather}
		\tilde{S}_{\text{G}}=\int\dd[2]{x}\dfrac{1}{2\la'}\tensor{\e}{^{\m}^{\n}}\tensor{\e}{_{a}_{b}}(\tensor{e}{_{\m}^{a}}-\tensor{f}{_{\m}^{a}})(\tensor{e}{_{\n}^{b}}-\tensor{f}{_{\n}^{b}})\,, \label{tolgr}
\end{gather}
in agreement with \cite{Tolley:2019nmm}. Once again, it is important to stress that for quadratic operators which depend on \(\la\) and no other couplings, the initial condition \eqref{inco} seems to be consistent only when \(b_{2}=0\).
\subsubsection{Retrieving \(T\overline{T}+\Lambda_{2}\)}
Starting on the same footing as before, we pick the solution that obeys \eqref{con1} but now \(b_{2}=0\). Considering this, we will prove that at least one of the coefficients \(\mathfrak{c_{i}}\) carries units necessarily by contradiction. Let us assume that all \(\mathfrak{c_{i}}\) are unitless. Then, we once more have that \(\mathfrak{c_{0}}=0=\mathfrak{c_{2}}\) and \(b=1\) but now the condition for \(b_{3}\) is \(b_{3}=-1/(8b_{0})\). For these choices there does not exist any non-zero value\footnote{If \(\mathfrak{c_{1}}=0\), the deforming parameter completely decouples from the auxiliary zweibein, which is not compatible with our methodology.} of \(\mathfrak{c_{1}}\) such that \(G\) has the correct units.

From there, we understand that \(G\) will now be a function of other couplings as well, which may or may not be irrelevant \cite{Gorbenko:2018oov, Coleman:2021nor}. In order for us to get \(T\overline{T}+\Lambda_{2}\), we may now proceed with the following identification
\begin{gather}
	\begin{gathered}
	b_{0}\longleftrightarrow\frac{1}{2\al}\,, \qquad b_{3}\longleftrightarrow\al\frac{\eta-1}{4}\,,\\
	\mathfrak{c_{0}}(\la_0)\longleftrightarrow-\la_{0}^{\sqrt{\eta}}\,, \qquad \mathfrak{c_{1}}(\la_0)\longleftrightarrow-\al\sqrt{\eta}\sqrt{\la_{0}}^{\sqrt{\eta}-1}\,, \qquad \mathfrak{c_{2}}(\la_0)\longleftrightarrow\al\frac{\sqrt{\eta}-1}{4\la_{0}}\,,
\end{gathered}
\end{gather}
where \(\al\) and \(\eta\) are arbitrary constants. This leads to the following massive gravity action
\begin{gather}
	S_{\text{G}}=\int\dd[2]{x}\dfrac{\al}{2}\tensor{\e}{^{\m}^{\n}}\tensor{\e}{_{a}_{b}}\qty{\be_{-}(\la,\la_{0})\frac{\tensor{f}{_{\m}^{a}}\tensor{f}{_{\n}^{b}}}{2\la}+\be_{3}(\la,\la_{0})\frac{\tensor{f}{_{\m}^{a}}\tensor{e}{_{\n}^{b}}}{\sqrt{\la_{0}\la}}-\be_{+}(\la,\la_{0})\frac{\tensor{e}{_{\m}^{a}}\tensor{e}{_{\n}^{b}}}{2\la_{0}}}\,,
\end{gather}
with
\begin{gather}
	\be_{\pm}(\la,\la_{0})\coloneqq1\pm\sqrt{\eta}\coth(\frac{\sqrt{\eta}}{2}\log\frac{\la_{0}}{\la})\,, \quad	\be_{3}(\la,\la_{0})\coloneqq\sqrt{\eta}\csch(\frac{\sqrt{\eta}}{2}\log\frac{\la_{0}}{\la})\,,
\end{gather}
in agreement with \cite{Torroba:2022jrk}. For \eqref{2ddef}, the term that comes with \(b_{3}\) in front is the one that is analogous to \(\Lambda_{2}\). Setting \(\eta=1\), \(\al=1/(2b_{0})\) and \(\la_{0}=0\) in that order, gives \eqref{tolgr} as expected.
\subsection{Spectrum analysis} \label{secspec}
It is rather interesting to explore the energy spectrum of the operator \eqref{2ddef} when it is constrained only by \eqref{con1} and placed on an Euclidean cylinder of circumference \(R\)  \cite{Zamolodchikov:2004ce}. The details of this calculation are quite standard and are located in appendix \ref{derspec}. In short, we find that the energies are given by
\begin{gather}
	E^{\pm}_{n}(\la',R)=-\dfrac{R\qty(1+2b_{2})}{2\la'}\pm\sqrt{\qty(\dfrac{\mathcal{P}_{n}}{R})^{2}+\qty(\dfrac{R\qty(1+2b_{2})}{2\la'})^{2}+\dfrac{2b_{0}\qty(b_{3}R^{2}+\la'\mathcal{E}_{n})}{{\la'}^{2}}}\,, \label{solsprecmain}
\end{gather}
where \(\mathcal{E}_{n}\) are the undeformed energies that survive the \(\la\rightarrow 0\) limit and \(\mathcal{P}_{n}=2\pi p_{n}\), with \(p_{n}\in\mathds{Z}\). One notices that as in the standard \(T\overline{T}\) case, there are two branches for the energy and usually the divergent one gets discarded. We choose to keep both, as there are cases e.g. \cite{Iliesiu:2020zld, Rosso:2020wir,Griguolo:2021wgy} where both branches are necessary. Taking now the \(\la\rightarrow 0\) limit, one retrieves a non-divergent value only for \(b_{3}=0\) and \(b_{2}\neq-1/2\) i.e.
\begin{gather}
	E^{\pm}_{n}(R)=\frac{2b_{0}\mathcal{E}_{n}}{R(1+2b_{2})}\,, 
\end{gather}
where \(b_{2}>-1/2\) for the positive, and \(b_{2}<-1/2\) for the negative branch respectively. Keeping \(b_{3}\) at zero and defining \(R'\coloneqq R(1+2b_{2})\) and \(\mathcal{P}'_{n}\coloneqq\mathcal{P}_{n}(1+2b_{2})\) one gets back the standard \(T\overline{T}\) spectrum, but now the circumference of the cylinder and the momenta are scaled. Considering now the second constraint \eqref{con2}, we uniquely reach \(b_{2}=0\) --which is consistent with \eqref{tolgr}-- for the positive and \(b_{2}=-1\) for the negative branch.  

The spectrum assumes a simpler form when we consider \eqref{con2}, with the terms related to the coefficient \(b_{2}\) cancelling exactly inside the square root, yielding
\begin{gather}
	E^{\pm}_{n}(\la',R)=-\dfrac{R\qty(1+2b_{2})}{2\la'}\pm\sqrt{\qty(\dfrac{2\pi p_{n}}{R})^{2}+\qty(\dfrac{R}{2\la'})^{2}+\dfrac{2b_{0}\mathcal{E}_{n}}{\la'}}\,.
\end{gather}
This equation is tantalisingly reminiscent of the string spectrum in \cite{Danielsson:2000gi}, which matches with our result up to the following non-unique identifications\footnote{For a different interpretation see \cite{Blair:2020ops}.}
\begin{gather}
		\la'\leftrightarrow\dfrac{\al'\pi}{w}\,, \quad b_{2}\leftrightarrow\dfrac{\tensor{B}{_{0}_{1}}-1}{2}\,, \quad p_{n}\leftrightarrow\dfrac{N_{\text{L}}-N_{\text{R}}}{w}\,, \quad \mathcal{E}_{n}\leftrightarrow\dfrac{\pi\qty(N_\text{L}+N_{\text{R}}+k_{\bot}^{2})}{wb_{0}}\,,
\end{gather}
when \(p_{\bot}^{2}=2k_{\bot}^{2}/\al'\). Moreover, the \(B\)-field in similar fashion with \(b_{2}\), cancels exactly inside the square root and this is consistent with the identification above. The units also seem to match, yet \(b_{2}\) is constant in contrast to \(\tensor{B}{_{0}_{1}}\), which is a function of the worldsheet coordinates, therefore this identification is limited to cases with a constant \(B\)-field. Finally, turning off the \(B\)-field, which corresponds to \(b_{2}=-1/2\), we trivially retrieve a more familiar string theory spectrum \cite{Polchinski:1998rq}. 

It is important to note that we retrieve known string theory spectra for certain choices of \(b_{2}\) that do not necessarily have a finite \(\la\rightarrow 0\) limit. This is due to the identification above, essentially \(\la\) plays the role of \(\al'\) and consequently the string tension grows as the deformation parameter approaches zero. When we finally get there, the tension becomes infinite, followed by the energy and vice versa, in accordance with the derived spectrum. Regarding the analysis at the level of the action, this is a strong indicator towards the relaxation of the initial condition \eqref{inco}. Indeed, in subsection \ref{sigmodel} we reach that conclusion.           
\subsection{Correlators of undeformed operators} \label{correl}
Let us now shift our attention to correlators. For simplicity,\footnote{Technically, one should be able to begin this discussion from the general solution \eqref{2dsol}, potentially retrieving more properties regarding correlators. With our choice, we are essentially covering all deformations that depend explicitly only on \(\la\).} consider the complete solution \eqref{sol1}, the massive gravity action \eqref{soll2d} and consequently the deformed partition function i.e.
\begin{gather}
	\mathcal{Z}_{\la}=\int\mathcall{D}\Phi\hspace{0.02cm}\mathcall{D}e\hspace{0.1cm}e^{iS_{\text{G}}[\tensor{e}{_{\m}^{a}},\tensor{f}{_{\m}^{a}},\la]+iS_{0}[\tensor{e}{_{\m}^{a}},\mathfrak{g}_{i},\Phi]}=\int\mathcall{D}\Phi\hspace{0.1cm}e^{iS[\tensor{f}{_{\m}^{a}},\la,\mathfrak{g}_{i},\Phi]}\,, \label{path1}
\end{gather}
where the deforming operator is given by \eqref{opp2d} and assuming that the path integral over the auxiliary zweibein can be performed. Staring at equation \eqref{soll2d} for a bit leads to the following equality
\begin{gather}
	\int\mathcall{D}\Phi\hspace{0.02cm}\mathcall{D}e\hspace{0.1cm}\exp(i\int\dd[2]{x}\dfrac{f}{\la'}\hspace{0.02cm}b_{2}+i\tilde{S}_{\text{G}}+iS_{0})=\int\mathcall{D}\Phi\hspace{0.1cm}\exp(i\int\dd[2]{x}\dfrac{f}{\la'}\hspace{0.02cm}b_{2}+iS_{T\overline{T}})\,, \label{path2}
\end{gather}
where \(\tilde{S}_{\text{G}}\) is given by \eqref{tolgr}, i.e. the standard \(T\overline{T}\) massive gravity action and \(S_{T\overline{T}}\) is now the \(T\overline{T}\) deformed action. The role of \(b_{2}\) now becomes apparent, it behaves like a source term for the deformed theory. The only element missing is that \(b_{2}\) is constant and not an arbitrary function of spacetime. We can actually promote \(b_{2}\) to a function if we are careful in doing so. To make the argument clear, we briefly revisit equation \eqref{gsol} and \eqref{2dsol}, essentially the generic massive gravity solution to the operator \eqref{2ddef}. Now we boldly perform the following uplift
\begin{gather}
	b_{i}\longrightarrow b_{i}(x)\,, \qquad i\in\{0,1,2,3\}\,, \label{uplf}
\end{gather}
and check if this affects any part of our algorithm. Retracing all the steps we took in the beginning of this section, we notice that this uplift is always valid unless any of the \(b_{i}(x)\) is a function of \(\tensor{e}{_{\m}^{a}}\) or/and \(\tensor{f}{_{\m}^{a}}\) or/and \(\la\). That is because we essentially perform three types of variations throughout the process of finding the massive gravity solution: variations with respect to \(\tensor{f}{_{\m}^{a}}\) to calculate the deformed stress-tensor and consequently construct the deforming operator, variations with respect to \(\la\) for the flow equation and finally,  variations with respect to \(\tensor{e}{_{\m}^{a}}\) in order to eliminate the auxiliary zweibein. As we mentioned in section \ref{sec2} this last step is classical, but it also constraints the fully quantum mechanical approach in the same way. Following this, the source term can essentially be anything that does not couple to the background and does not depend on the deformation parameter. In other words, we are now able to calculate deformed correlators of undeformed operators.

Returning to \eqref{path1}, we now set \(b_{2}(x)=J(x)\mathcal{O}(x)\) (to be more general, we assume that \(b_{0}(x)\) is also a function) and define the \(T\overline{T}\) deformed correlators as follows
\begin{gather}
	\expval{\mathcal{O}(x_{1})\cdots\mathcal{O}(x_{n})}_{\la}\coloneqq\dfrac{\la'(x_{1})\cdots\la'(x_{n})}{i^{n}f(x_{1})\cdots f(x_{n})}\dfrac{1}{\mathcal{Z}_{\la}[J]}\dfrac{\de\mathcal{Z}_{\la}[J]}{\de J(x_{1})\cdots\de J(x_{n})}\Bigg{|}_{J=0}\,. \label{corr1}
\end{gather}
Furthermore, we can actually relate the deformed and undeformed correlators using \eqref{path2}, which takes the following neat form
\begin{gather}
	\mathcal{Z}_{\la}[\tensor{f}{_{\m}^{a}}]\expval{\mathcal{O}(x_{1})\cdots\mathcal{O}(x_{n})}_{\la}=\int\mathcall{D}e\hspace{0.1cm}e^{i\tilde{S}_{\text{G}}[\tensor{e}{_{\m}^{a}},\tensor{f}{_{\m}^{a}},\la]}\mathcal{Z}_{0}[\tensor{e}{_{\m}^{a}}]\expval{\mathcal{O}(x_{1})\cdots\mathcal{O}(x_{n})}_{0}\,, \label{corr2}
\end{gather}
where \(\mathcal{Z}_{0}\) is the undeformed partition function and the sources have been turned off. It is quite interesting to observe that the equation above looks like an integral transformation where the kernel is the exponential of \(i\tilde{S}_{\text{G}}\). At \(\la\rightarrow 0\) the kernel essentially becomes a delta functional with support only at  \(\tensor{e}{_{\m}^{a}}=\tensor{f}{_{\m}^{a}}\) and consequently the equality holds trivially as expected. Moreover, setting all operators to unity we recover the formula that relates the \(T\overline{T}\)-deformed partition function with the undeformed one \cite{Tolley:2019nmm}. A comparison between our results and previous treatments \cite{Kraus:2018xrn,Cardy:2019qao,Cui:2023jrb,Aharony:2023dod,Barel:2024dgv} would be interesting.
\subsection{Dimensional reduction, dilaton gravity theories and point particles}\label{secdimred}
Another key aspect of \(T\overline{T}\) is the well-known trace-flow equation, that relates the trace of the deformed stress-tensor with its determinant classically. The combination of (\ref{wa}, \ref{eqmot}) and \eqref{soll2d} leads to the following, more general, trace-flow equation
\begin{gather}
	\dfrac{\bar{e}\tr\bar{\mathcall{S}}}{2}=\dfrac{f\tr\bar{T}}{2}-\la'f\det\tensor{\bar{T}}{^{\mu}_{\nu}}+fb_{2}\tr\bar{T}-\dfrac{fb_{2}(1+b_{2})}{\la'}\,, \label{trflow}
\end{gather}
which will be the main actor of this subsection. We will always assume that we are on-shell for the auxiliary zweibein, thus we can safely drop the ``bar'' notation. In \cite{Gross:2019ach}, the authors begin from the standard \(T\overline{T}\) description and the corresponding flow equation for a seed conformal field theory, and dimensionally reduce via circular compactification of the spacial dimension with circumference set to one, followed by the trivial elimination of the anti-diagonal elements of the stress-tensor.

We will now repeat the same process, but for \eqref{opp2d} and \eqref{trflow}. The first step is to eliminate the spacial component using the trace-flow equation, thus we begin by picking our coordinates to be \((t,\theta)\). Then, solving the trace-flow equation for \(\tensor{T}{^{\theta}_{\theta}}\) gives
\begin{gather}
\tensor{T}{^{\theta}_{\theta}}=\dfrac{2b_{2}(1+b_{2})-\tensor{T}{^{t}_{t}}(1+2b_{2})\la'-2\tensor{T}{^{t}_{\theta}}\tensor{T}{^{\theta}_{t}}{\la'}^{2}}{(1+2b_{2}-2\tensor{T}{^{t}_{t}}\la')\la'}\,,	
\end{gather}
therefore \eqref{opp2d} becomes
\begin{gather}
	\mathcal{O}=\int\dd[2]{x}f\hspace{0.05cm}2b_{0}\dfrac{(\tensor{T}{^{t}_{t}}\!)^{2}+\tensor{T}{^{t}_{\theta}}\tensor{T}{^{\theta}_{t}}-b_{2}(1+b_{2})/{\la'}^{2}}{1+2b_{2}-2\tensor{T}{^{t}_{t}}\la'}\,. \label{2dop1d}
\end{gather}
We now proceed with the dimensional reduction by setting \(\tensor{T}{^{t}_{\theta}}=0=\tensor{T}{^{\theta}_{t}}\) and we Wick rotate \((\tau=it)\) to match the notation of \cite{Gross:2019ach}. Combining everything, the flow equation \eqref{floweq} takes the following form
\begin{gather}
	\pdv{S_{\text{E}}}{\la'}=\int\dd{\tau}\sqrt{\gamma}\hspace{0.1cm}\dfrac{T^{2}-b_{2}(1+b_{2})/{\la'}^{2}}{1+2b_{2}-2T\la'}\,, \label{1dred}
\end{gather}
where \(T\coloneqq\tensor{T}{^{\tau}_{\tau}}\), assuming that the metric determinant has no spacial dependence. The deformation above was derived using the Wheeler-DeWitt method \cite{Gross:2019ach,Hartman:2018tkw} and it is instructive to compare the two approaches. Briefly, one begins by considering the following dilaton gravity theory
\begin{gather}
	S_{\text{E}}=-\dfrac{1}{2\kappa^{2}}\int_{M}\dd[2]{x}\sqrt{g}\hspace{0.1cm}\{\Phi R+2U(\Phi)\}-\dfrac{1}{\kappa^{2}}\int_{\partial M}\dd{\tau}\sqrt{\tensor{g}{_{\tau}_{\tau}}}\hspace{0.1cm}\Phi\{K-1\}\,,
\end{gather}
where \(\tensor{g}{_{\tau}_{\tau}}\) is the boundary metric and \(K\) measures the extrinsic curvature. Now, using the methodology developed in \cite{Hartman:2018tkw}, the authors of \cite{Gross:2019ach} derive the following flow equation
\begin{gather}
	\pdv{S_{\text{EFT}}}{\la}=\int\dd{\tau}\sqrt{\gamma}\hspace{0.1cm}\dfrac{T^{2}-(1-(r_{c}\Phi_r)^{-1}U(r_{c}\Phi_r))/(4\la)^{2}}{1/2-2\la T}\,, \label{1ddir}
\end{gather}
where \(\Phi_r\coloneqq \Phi/r\) and \(r_{c}\) is the finite radial cut-off which is a function of \(\la\). 

Assuming now that this flow equation holds even when \(r_{c}\) is independent of \(\la\) and carefully comparing \eqref{1dred} with \eqref{1ddir}, we are lead to a yet another constraint, and an interesting identification, i.e.
\begin{gather}
	b_{2}=\dfrac{1}{2}(b_{0}-1)\,, \qquad 	b_{0}\longleftrightarrow\pm\sqrt{\dfrac{r_{c}\Phi_r}{U(r_{c}\Phi_r)}}\,. \label{matchhh}
\end{gather}
To be more precise, in order for the identification to be valid in general, one is obligated to perform exactly the same uplift as in \eqref{uplf}.  Since \(U(x)\) is a potential term and because we assumed that \(r_{c}\) does not depend on the deformation parameter, the criteria for the uplift are precisely satisfied. From there it follows that for the \(\text{AdS}_{2}\) potential, that is \(U(x)=x\) (where the assumption for the radial cut-off is redundant), we trivially retrieve \(b_{0}=1\) and \(b_{2}=0=b_{3}\) which reduces the operator \eqref{opp2d} to \(T\overline{T}\) as expected, as well as \(b_{0}=-1=b_{2}\) and \(b_{3}=0\) which again reduces the operator to \(T\overline{T}\) for seed conformal field theories using \eqref{trflow}. Finally, it is worthwhile to point out that there does exist a generalization of \eqref{1ddir} assuming additional matter to be present \cite{Gross:2019ach}. Our extension \eqref{1dred} cannot match the extra matter terms that appear, meaning that, potentially, one needs to begin from the more general solution \eqref{2dsol} where \(b_{2}\) and \(b_{3}\) are independent. Then, all \(\mathfrak{c_{i}}\) have to be fixed by introducing the dimensionful couplings \(\mathfrak{g}_{i}\) indicating the existence of new matter terms, in agreement with the aforementioned generalization. For the next step, the new trace-flow equation must be found via \eqref{wa} and \eqref{eqmot} with the dimensional reduction following in exactly the same way as before.\footnote{Please do note that since more couplings are now present, the trace of the undeformed stress-tensor may not be zero, which is exactly the source of new matter.} Lastly, the two approaches are compared side by side, yielding constraints and identifications similar to the ones above. It would be interesting to investigate under which conditions for \(b_{0}\) and \(b_{2}\), this identification holds when  \(r_{c}\) is a function of \(\la\). 

Let us now study the flow equation \eqref{1dred}. The Lorentzian version of the deforming operator reads 
\begin{gather}
	\mathcal{O}=\int\dd{t}f\hspace{0.05cm}2b_{0}\dfrac{T^{2}-b_{2}(1+b_{2})/{\la'}^{2}}{1+2b_{2}-2T\la'}\,, \label{1dopp}
\end{gather}
where \(T\coloneqq\tensor{T}{^{t}_{t}}\). Given this deformation, we can derive the corresponding massive gravity action and deform a theory classically. Since we are now living in one dimension, both einbeins are equal to their determinants. Similarly to \eqref{gsol}, we now proceed with a slightly different ansatz
\begin{gather}
	G(\la',\mathfrak{g}_{i},y_{1})=\dfrac{c_{0}(\la',\mathfrak{g}_{i})}{y_{1}}+c_{1}(\la',\mathfrak{g}_{i})+c_{2}(\la',\mathfrak{g}_{i})y_{1}\,,
\end{gather}
therefore, from \eqref{pde} and \eqref{1dopp} we retrieve the following massive gravity solution
\begin{gather}
S_{\text{G}}=\int\dd{t}\dfrac{f}{\la'}\qty{\dfrac{1}{2}+b_{2}-\dfrac{\mathfrak{c_{0}}(\mathfrak{g}_{i})}{4y_{1}}-\dfrac{y_{1}}{4\mathfrak{c_{0}}(\mathfrak{g}_{i})}+\la'y_{1}\mathfrak{c_{2}}(\mathfrak{g}_{i})}\,, \label{gr1d}
\end{gather}
where the units of \(\mathfrak{c_{i}}\) are
\begin{gather}
	\qty[\mathfrak{c_{0}}(\mathfrak{g}_{i})]=M^{0}\,,\qquad \qty[\mathfrak{c_{2}}(\mathfrak{g}_{i})]=M^{1}\,.
\end{gather}
The last ingredient that we will need is the seed function, which we choose to be a non-linear sigma model of bosons and fermions interacting with an arbitrary potential \(V\) plus topological terms which are denoted by \(\mathcal{T}\). The explicit action reads
\begin{gather}
	S_{0}=\int\dd{t}e\hspace{0.05cm}\qty{\dfrac{1}{2e^{2}}\tensor{A}{_{i}_{j}}\dot{\phi}^{i}\dot{\phi}^{j}\mp\dfrac{i}{2e}\tensor{B}{_{i}_{j}}\qty(\bar{\psi}^{i}\dot{\psi}^{j}-\dot{\bar{\psi}}^{i}\psi^{j})-V+\dfrac{\mathcal{T}}{e}}\,,
\end{gather}
where \(i,j\) are the flavour indices. There are two points in the equation above that we need to stress. First, it is clear that fermions behave exactly like the topological term in terms of their interaction with the background, in other words they decouple completely. Second, we understand that in reality we are not necessarily concerned about the exact structure of the matter fields, but only about the way they couple to the background. To make things more precise, consider the following definitions
\begin{gather}
	\Phi\coloneqq\tensor{A}{_{i}_{j}}\dot{\phi}^{i}\dot{\phi}^{j}\,, \qquad \Psi\coloneqq\pm i\tensor{B}{_{i}_{j}}\qty(\bar{\psi}^{i}\dot{\psi}^{j}-\dot{\bar{\psi}}^{i}\psi^{j})\,,
\end{gather}
which brings the seed action to a more convenient form
\begin{gather}
	S_{0}=\int\dd{t}e\hspace{0.05cm}\qty{\dfrac{1}{2e^{2}}\Phi-\dfrac{1}{2e}\qty(\Psi-2\mathcal{T})-V}\,. \label{seed1d}
\end{gather}
This simple rewriting essentially allows one to deform not just bosons and fermions but any matter action that couples to the background in the same way. This is one of the key advantages of the massive gravity method and the main reason for which these type of deformations are usually referred as purely ``geometrical''. We now combine (\ref{gr1d}, \ref{seed1d}) and solve the equation of motion for \(e\) which gives
\begin{gather}
	\bar{e}_{\pm}=\pm f\sqrt{\dfrac{\mathfrak{c_{0}}\qty(\mathfrak{c_{0}}-2\la'\Phi/f^{2})}{1+4\la'\mathfrak{c_{0}}(V-\mathfrak{c_{2}})}}\,,\label{sole1d}
\end{gather}
with the deformed action following
\begin{gather}
	S_{\mp}=\int\dd{t}f\qty{\dfrac{1+2b_{2}\mp\sqrt{\qty[1-2\la'\Phi/(\mathfrak{c_{0}}f^{2})]\qty[1+4\la'\mathfrak{c_{0}}(V-\mathfrak{c_{2}})]}}{2\la'}-\dfrac{1}{2f}\qty(\Psi-2\mathcal{T})}.
\end{gather}
Now applying the initial condition \eqref{inco} gives \(\mathfrak{c_{0}}=1\), \(\mathfrak{c_{2}}=0\) and \(b_{2}=0\), and for these choices, the theory above matches exactly the one given in \cite{Gross:2019ach}. 

Finally, the action above can be brought in the form of a relativistic particle for the static gauge choice when the fermions and the topological term are turned off. In order to do this we define the proper time and a target space metric as 
\begin{gather}
	\mathfrak{t}\coloneqq\int_{0}^{t}f(\tilde{t})\dd{\tilde{t}}\,, \qquad \tensor{h}{_{\m}_{\nu}}\coloneqq\tensor{h}{_{0}_{0}}\oplus\tensor{h}{_{i}_{j}}\,,
\end{gather}
and proceed with the following identification
\begin{gather}
	m^{-1}\longleftrightarrow2\la'\,,\quad \dot{X}^{i}\longleftrightarrow\sqrt{2\la'}\dot{\phi}^{i}\,,\quad \tensor{h}{_{\m}_{\nu}}\longleftrightarrow\qty[1+4\la'\mathfrak{c_{0}}(V-\mathfrak{c_{2}})]\qty[-1\oplus\tensor{A}{_{i}_{j}}/\mathfrak{c_{0}}]\,,
\end{gather}
where \(m\) is the mass of the particle and the derivatives are now with respect to proper time. This leads to the following action
\begin{gather}
	S_{\mp}=\mp\int\dd{\mathfrak{t}}m\sqrt{- \dot{X}^{\mu}\dot{X}^{\nu}\tensor{h}{_{\m}_{\nu}}}\,, \qquad \dot{X}^{0}=1\,,
\end{gather}
for \(b_{2}=-1/2\). We should note that the mass can be taken out of the integral only when \(b_{0}\) is not a function of proper time. During the next subsection we will repeat the same process but in two dimensions.
\subsection{Sigma models} \label{sigmodel}
After the one-dimensional detour, we are back to two dimensions and we would like to apply the massive gravity method to the same model as before, i.e. a non-linear sigma model of bosons and fermions interacting with an arbitrary potential plus topological terms. This has been done multiple times in the past using different approaches \cite{Cavaglia:2016oda,Kraus:2018xrn,Bonelli:2018kik,Frolov:2019nrr,Coleman:2019dvf,Caputa:2020lpa}, but we aim to provide a more transparent expression for the deformed action. Let us begin from the seed action which reads
\begin{gather}
	S_{0}=\int\dd[2]{x}e\hspace{0.05cm}\qty{-\dfrac{1}{2}\tensor{g}{^{\mu}^{\nu}}\tensor{\Phi}{_{\mu}_{\nu}}-\dfrac{1}{2}\tensor{e}{^{\mu}_{a}}\tensor{\Psi}{^{a}_{\mu}}-V+\dfrac{\mathcal{T}}{e}}\,, \label{seed2d}
\end{gather}
where \(\tensor{\Phi}{_{\mu}_{\nu}}\) and \(\tensor{\Psi}{^{a}_{\mu}}\) have similar definitions as before i.e.
\begin{gather}
\tensor{\Phi}{_{\mu}_{\nu}}\coloneqq\tensor{A}{_{i}_{j}}\partial_{(\mu}\phi^{i}\partial_{\nu)}\phi^{j}\,, \qquad \tensor{\Psi}{^{a}_{\mu}}\coloneqq\tensor{B}{_{i}_{j}}\bar{\psi}^{i}\g^{a}\vecev{\nabla}_{\mu}\psi^{j}=\tensor{B}{_{i}_{j}}\bar{\psi}^{i}\g^{a}\vecev{\partial}_{\mu}\psi^{j}\,.
\end{gather}
Once again, this rewriting allows one to deform any theory that couples in the same way to the background. Proceeding now with the gravity action \eqref{soll2d} and the seed action above, we find the following solution for the auxiliary zweibein
\begin{gather}
	\tensor{{(\bar{e}_{\pm})}}{_{\mu}^{a}}=\dfrac{\la'}{2\widetilde{\la'}}\qty(\tensor{\widehat{f}}{_{\mu}^{a}}\pm\dfrac{\tensor{\widehat{f}}{_{\mu}^{a}}+2\widetilde{\la'}\tensor{\Phi}{_{\mu}_{\nu}}\tensor{\widehat{f}}{^{\nu}_{b}}\tensor{\eta}{^{b}^{a}}}{\sqrt{\widehat{\g}^{-1}\det(\tensor{\widehat{\g}}{_{\mu}_{\nu}}+2\widetilde{\la'}\tensor{\Phi}{_{\mu}_{\nu}})}})\,, \label{2dzweisol}
\end{gather}
where
\begin{gather}
	\widetilde{\la'}\coloneqq\la'(1-V\la')\,, \qquad \tensor{\widehat{f}}{_{\mu}^{a}}\coloneqq\tensor{f}{_{\mu}^{a}}+\dfrac{\la'}{2}\tensor{\Psi}{_{\mu}^{a}}\,,\qquad \tensor{\widehat{\g}}{_{\mu}_{\nu}}\coloneqq\tensor{\widehat{f}}{_{\mu}^{a}}\tensor{\eta}{_{a}_{b}}\tensor{\widehat{f}}{_{\nu}^{b}}\,.
\end{gather}
For the case of a single scalar boson, without fermions, the solution above reduces to the one found in \cite{Caputa:2020lpa}. Using \eqref{2dzweisol} we obtain the following deformed action
\begin{gather}
	S_{\pm}=\int\dd[2]{x}f\hspace{0.05cm}\qty{\dfrac{1+b_{2}}{\la'}-\dfrac{1}{2\widetilde{\la'}f}\qty(\sqrt{-\det(\tensor{\widehat{\g}}{_{\mu}_{\nu}})}\pm\sqrt{-\det(\tensor{\widehat{\g}}{_{\mu}_{\nu}}+2\widetilde{\la'}\tensor{\Phi}{_{\mu}_{\nu}})})+\dfrac{\mathcal{T}}{f}}\,. \label{sol2dlin}
\end{gather}
There are certain comments that need to be made regarding the action above. First, the expression agrees with \eqref{solsigma}, which was derived using the ansatz method with the fermions turned off. Furthermore, for \(b_{2}=0\) the action agrees with the ones derived in \cite{Frolov:2019nrr,Coleman:2019dvf} and consequently with \cite{Cavaglia:2016oda,Kraus:2018xrn,Bonelli:2018kik,Caputa:2020lpa}, hence its validity is secured. Moreover we observe something interesting when the bosons and potential are absent. The solution \eqref{2dzweisol} now reads
\begin{gather}
	\tensor{{(\bar{e}_{+})}}{_{\mu}^{a}}=\tensor{f}{_{\mu}^{a}}+\dfrac{\la'}{2}\tensor{\Psi}{_{\mu}^{a}}\,,
\end{gather}
which is linear in \(\la'\). This is important because in that limit, the deformed action is also linear in \(\la'\). Considering these facts, we naturally conclude that the source of this linear dependence comes entirely from the way fermions couple to the background and not due to their anti-commuting nature. This is consistent with the fact that \(\tensor{\Psi}{^{a}_{\mu}}\) may represent any matter field with this index structure. In other words, visiting \eqref{seed2d}, one notices that \(\tensor{\Psi}{^{a}_{\mu}}\) couples linearly to the background, therefore the deformed action will also be linear but now in \(\la'\). This answers a question posed by the authors of \cite{Coleman:2019dvf}.

Finally, we observe that the square root structures above for a flat background, can be interpreted as Nambu-Goto densities up to certain identifications for the target space coordinates and string tension \cite{Cavaglia:2016oda}, with \(\tensor{\widehat{\g}}{_{\mu}_{\nu}}\) and \(\tensor{\widehat{\g}}{_{\mu}_{\nu}}+2\widetilde{\la'}\tensor{\Phi}{_{\mu}_{\nu}}\) being the world-sheet metrics respectively. Turning off the potential and topological terms makes this identification more precise in a way that it is now consistent with the spectrum analysis in subsection \ref{secspec}. In more detail, the choice \(b_{2}=-1/2\) once again kills the constant term in \eqref{sol2dlin} as expected.
\subsection{Scalar boson of order \(k\)}
Given our results, we would like to explore how higher-derivative theories are deformed. For simplicity, consider the conformal scalar boson of order \(k\) as a seed theory, temporarily in \(d\) dimensions. Our methodology is valid only for theories that couple to a curved background, which in our case translates to
\begin{gather}
	S_{0}=\dfrac{1}{2}\int\dd[d]{x}\phi\Box^{k}\phi \longrightarrow\dfrac{1}{2}\int\dd[d]{x}e\hspace{0.05cm}\qty{\phi\Box^{k}\phi+\text{curvature terms}}\,,
\end{gather} 
where the curvature terms can be fixed exactly for any \(k\in \mathds{N}^{+}\) \cite{ojm/1200689814,gsmsv1,10.1112/jlms/s2-46.3.566,Paneitz_2008}. There are two issues with this approach. First, curvature couplings are introduced, therefore the second derivative of the auxiliary vielbein will appear, which in turn means that finding the vielbein solution to the equations of motion for \(S_{\text{G}}+S_{0}\) is improbable. The second and more serious problem arises due to the dependence of the curvature terms on \(d\). For example, at \(k=2\) there are curvature terms that carry coefficients which are proportional to \((d-2)^{n}\) where \(n\) is negative \cite{Brust:2016gjy}, hence one cannot simply set \(d=2\). 

To avoid these issues entirely we will attack the problem from a different angle. Let us return to two dimensions where we will introduce \((k-1)\) auxiliary scalar fields \cite{Bergshoeff:2012sc,Gibbons:2019lmj} and consider the following action
\begin{gather}
S_{0}=\int\dd[2]{x}e\hspace{0.05cm}\qty{-\dfrac{1}{2}\tensor{g}{^{\mu}^{\nu}}\tensor{A}{_{i}_{j}}\partial_{(\mu}\phi^{i}\partial_{\nu)}\phi^{j}-\dfrac{1}{2}\tensor{B}{_{i}_{j}}\phi^{i}\phi^{j}-\mathcal{V}(\phi^{k},\mathfrak{g}_{i})}\,,\qquad 1\leq i,j \leq k\,,	\label{ksigma}
\end{gather}  
where the matrices \(A\) and \(B\) are given by
\begin{gather}
	\tensor{A}{_i_j}=\begin{pmatrix}0&\cdots&0&1\\ \vdots&\rotatebox{10}{\(\iddots\)}&\rotatebox{10}{\(\iddots\)}&0\\0&\rotatebox{10}{\(\iddots\)}&\rotatebox{10}{\(\iddots\)}&\vdots\\1&0&\cdots&0\end{pmatrix}\,, \qquad \tensor{B}{_i_j}=\begin{pmatrix}0&\cdots&0&1&\omega\\ \vdots&\rotatebox{10}{\(\iddots\)}&\rotatebox{10}{\(\iddots\)}&\rotatebox{10}{\(\iddots\)}&0\\0&\rotatebox{10}{\(\iddots\)}&\rotatebox{10}{\(\iddots\)}&\rotatebox{10}{\(\iddots\)}&\vdots\\1&\rotatebox{10}{\(\iddots\)}&\rotatebox{10}{\(\iddots\)}&\rotatebox{10}{\(\iddots\)}&\vdots\\\omega&0&\cdots&\cdots&0\end{pmatrix}\,,
\end{gather}
with \(\omega\) being an arbitrary dimensionless constant. The equations of motion for this model are \cite{Bergshoeff:2012sc,Gibbons:2019lmj}
\begin{gather}
	\qty(\Box-\omega)\phi^{1}=\pdv{\mathcal{V}}{\phi^{k}}\,, \qquad \qty(\Box-\omega)\phi^{i}=\phi^{i-1}\,, \qquad 2\leq i \leq k\,,
\end{gather}
thus, using these we can eliminate the first \((k-1)\) fields, leaving us with an action that depends only on \(\phi^{k}\), which we will rename to \(\phi\) for simplicity. Then, for \(\omega=0\) we reach
\begin{gather}
S_{0}=\int\dd[2]{x}e\hspace{0.05cm}\qty{\dfrac{1}{2}\phi\Box^{k}\phi-\mathcal{V}(\phi,\mathfrak{g}_{i})}\,.	
\end{gather}
One notices that essentially, the role of the auxiliary fields is to maximally reduce the order of the derivatives. In our case, the reduced action is a simple sigma model plus a potential, which can be deformed easily using \eqref{sol2dlin} and setting  
\begin{gather}
	V=-\dfrac{1}{2}\tensor{B}{_{i}_{j}}\phi^{i}\phi^{j}-\mathcal{V}\,,
\end{gather}
where fermions and topological terms have been turned off. From there, one has two choices: either attempt to integrate out the auxiliary scalar fields assuming that the two processes commute, or work exclusively with \eqref{ksigma} where the number of derivatives is under control. We propose that the latter may be more productive, as any attempt to integrate out the auxiliary fields, even in the simplest case, did not produce any useful outcome. We should also note that the addition of topological terms might no longer be trivial since they may depend on the scalar fields, which in turn will affect the equations of motion. Finally, this approach may not be unique only to scalar fields and it would be interesting to study the generalisation to fields with higher spin.      
\subsection{Root deformations} \label{secroot}
The final stop in this two-dimensional journey is the study of the root-\(T\overline{T}\) deformation from the massive gravity perspective. The marginal operator that generates this deformation is given by \cite{Ferko:2022cix}
\begin{gather}
	\mathcal{R}=\int\dd[2]{x}f\hspace{0.05cm}\dfrac{b_{\text{r}}}{\sqrt{2}}\sqrt{\tensor{T}{^\mu_\nu}\tensor{T}{^\nu_\mu}-\frac{1}{2}(\tensor{T}{^\mu_\mu})^{2}}\,, \label{ttlikemain} 
\end{gather}
where \(b_{\text{r}}\) is a dimensionless constant. We will attach the letter \(\mu\) to the deformation parameter for the operator above. Proceeding as usual, equation \eqref{pde} for this choice becomes
\begin{gather}
\partial_{\mu}G=b_{\text{r}}\sqrt{2y_{2}-y_{1}^{2}}(\partial_{y_{1}}G+2y_{1}\partial_{y_{2}}G)\,, \label{ttlikepdemain}
\end{gather}
where \(b_{\text{r}}\) has absorbed a \(\pm\) factor, that is equation \eqref{ttlikepde} precisely. This is interesting. If we consider the approach we followed in subsection \ref{sigan}, the flow equation for the deformed Lagrangian density was derived via a sigma model seed theory ansatz which is entirely constructed via scalar fields. In contrast, the same flow equation is derived for \(G\), which is also a Lagrangian density, but now the building blocks are zweibeins and even more so, we have not specified the seed theory. This poses as a strong indicator towards the fact that root-\(T\overline{T}\) deformations are universal and independent of background-matter couplings, which in turn means that they cannot have a massive gravity description, at least not in the same way as the deformations that admit one. We can actually provide a more concrete argument about this fact, by solving the equation above. Let us temporarily forget that the solutions can be brought in the form of (\ref{ttlikesol1}, \ref{ttlikesol2}) and solve \eqref{ttlikepdemain} from scratch. We find
\begin{gather}
	G(\mu,\mathfrak{g}_{i},y_{1},y_{2})=\mathcall{G}\qty[\mathfrak{g}_{i},y_{1}^{2}-y_{2},\mu b_{\text{r}}+\arctanh\qty(y_{1}^{-1}\sqrt{2y_{2}-y_{1}^{2}})]\,, \label{ttlikemainsol}
\end{gather}
where \(\mathcall{G}\) is any function. We immediately notice that the deformation parameter decouples completely from \(y_{1}\) and \(y_{2}\) in the arguments of the solution. Moreover, applying the initial condition \eqref{inco} with \(\mu\rightarrow 0\) does not uniquely fix \(G\), yet a choice that is consistent reads
\begin{gather}
	G(\mu,\mathfrak{g}_{i},y_{1},y_{2})=\mathcall{F}(\mathfrak{g}_{i})\qty(y_{1}^{2}-y_{2}+\mu b_{\text{r}}+\arctanh\qty(y_{1}^{-1}\sqrt{2y_{2}-y_{1}^{2}})-2)\,,
\end{gather}   
where \(\mathcall{F}\) is an arbitrary dimensionless function. Since any solution that satisfies \eqref{inco} should provide a massive gravity description and a counterexample exists, we conclude that \eqref{ttlikemain} cannot have a massive gravity description on its own. This behaviour is reminiscent of gauge symmetry, in the way that it should work for any gauge choice. Nevertheless, the alternative interpretation is that the initial condition suggests that a massive gravity density is already present, possibly due to other deformations, and the effect of the root-\(T\overline{T}\) is the additional structure given by the solution above. Considering this, let us now write down the more useful form of the solution \cite{Ferko:2022cix} i.e. (\ref{ttlikesol1}, \ref{ttlikesol2}) or
\begin{gather}
	G(\mu,\mathfrak{g}_{i},y_{1},y_{2})=\mathcall{G}\qty(\mathfrak{g}_{i},y_{1,\m},y_{2,\m})\,, \label{ttlikesol1main}
\end{gather}
where
\begin{gather}
	\begin{aligned}
		&y_{1,\m}=y_{1}\cosh(b_{\text{r}}\m)+\sqrt{2y_{2}-y_{1}^{2}}\sinh(b_{\text{r}}\m)\,,\\&y_{2,\m}=y_{2}\cosh(2b_{\text{r}}\m)+y_{1}\sqrt{2y_{2}-y_{1}^{2}}\sinh(2b_{\text{r}}\m)\,,
	\end{aligned} \label{ttlikesol2main}
\end{gather}
and the initial condition has been arbitrarily (but very conveniently) set at \(\m=0\), where \(\mathcall{G}\qty(\mathfrak{g}_{i},y_{1,\m},y_{2,\m}){|}_{\m=0}\) reduces to the ``seed'' gravity action \(\mathcall{G}\qty(\mathfrak{g}_{i},y_{1},y_{2})\). Interestingly, the second argument in \eqref{ttlikemainsol} indicates that \(y_{2,\m}-y_{1,\m}^{2}=y_{2}-y_{1}^{2}\), which is verified by the set of equations above.

Combining all of the above we understand that in order to deform exclusively with root-\(T\overline{T}\) we have to follow a very specific procedure. First, we need to turn on a deformation, say with \(\la\) as a coupling, that admits a massive gravity solution. Then, using equations \eqref{ttlikesol1main} and \eqref{ttlikesol2main}, we introduce the root deformation as discussed and integrate out the auxiliary zweibein in order to ``push'' both deformations on the seed theory. Finally, we need to turn off \(\la\) to remove the effect of the extra deformation. Actually, since the root-\(T\overline{T}\) structure is unaffected by the ``seed'' massive gravity, the choice one makes originally for a deformation does not matter. We can actually take things a step further by taking advantage of that fact, and provide some more insight about root-\(T\overline{T}\).

Let us consider the solution \eqref{ttlikesol1main} and a seed action as usual. Then their combination gives the deformed theory \eqref{osol2}
\begin{gather}
	S[\tensor{e}{_{\m}^{a}},\tensor{f}{_{\m}^{a}},\la,\mu,\mathfrak{g}_{i},\Phi]=\int\dd[2]{x}f\hspace{0.05cm}\mathcall{G}\qty(\la,\mathfrak{g}_{i},y_{1,\m},y_{2,\m})+S_{0}[\tensor{e}{_{\m}^{a}},\mathfrak{g}_{i},\Phi]\,. \label{root2dsol}
\end{gather}
We now define the zweibein \(\tensor{\varepsilon}{_{\mu}^{a}}\) such that
\begin{gather}
	y_{1}^{\star}\coloneqq\tensor{f}{^{\nu}_{b}}\tensor{\varepsilon}{_{\nu}^{b}}=y_{1,\mu}\,,\qquad y_{2}^{\star}\coloneqq\tensor{f}{^{\nu}_{b}}\tensor{\varepsilon}{_{\la}^{b}}\tensor{f}{^{\la}_{c}}\tensor{\varepsilon}{_{\nu}^{c}}=y_{2,\m} \,, \label{defroot}
\end{gather}
where the \(\mu\) dependence is now implicit. In order to rewrite everything in terms of this new auxiliary zweibein, we combine \eqref{ttlikesol2main} and \eqref{defroot} to get
\begin{gather}
	\tensor{\varepsilon}{_{\n}^{a}}=\tensor{e}{_{\n}^{a}}\cosh(b_{\text{r}}\m)+\dfrac{2\tensor{e}{_{\n}^{b}}\tensor{f}{^{\la}_{b}}\tensor{e}{_{\la}^{a}}-y_{1}\tensor{e}{_{\n}^{a}}}{\sqrt{2y_{2}-y_{1}^{2}}}\sinh(b_{\text{r}}\m)\,. \label{root2dsolfin1}
\end{gather}
 Also, one notices that inverting \eqref{ttlikesol2main} leads to the exact same set of equations, but now \(y_{1}\leftrightarrow y_{1,\mu}\) and \(y_{2}\leftrightarrow y_{2,\mu}\). From there,
\begin{gather}			   \tensor{e}{_{\n}^{a}}=\tensor{\varepsilon}{_{\n}^{a}}\cosh(b_{\text{r}}\m)+\dfrac{2\tensor{\varepsilon}{_{\n}^{b}}\tensor{f}{^{\la}_{b}}\tensor{\varepsilon}{_{\la}^{a}}-y_{1}^{\star}\tensor{\varepsilon}{_{\n}^{a}}}{\sqrt{2y_{2}^{\star}-(y_{1}^{\star})^{2}}}\sinh(b_{\text{r}}\m)\,, \label{root2dsolfin}
\end{gather}
where \eqref{defroot} was used to avoid inconsistencies. Using \eqref{root2dsolfin}, \eqref{root2dsol} becomes
\begin{gather}
	S[\tensor{\varepsilon}{_{\m}^{a}},\tensor{f}{_{\m}^{a}},\la,\mu,\mathfrak{g}_{i},\Phi]=\int\dd[2]{x}f\hspace{0.05cm}\mathcall{G}\qty(\la,\mathfrak{g}_{i}, y_{1}^{\star},y_{2}^{\star})+S_{0}[\tensor{\varepsilon}{_{\m}^{a}},\mu,\mathfrak{g}_{i},\Phi]\,. \label{rename2droot}
\end{gather}
A direct comparison between \eqref{root2dsol} and \eqref{rename2droot} shows that the root-\(T\overline{T}\) deformation is ``pushed'' to the seed theory, which now lives on the background given by \eqref{root2dsolfin}. This essentially means that, upon correct identification of the matter Lorentz scalars \(x_{1}\) and \(x_{2}\) in the seed theory \eqref{eq7}, the uplift is \eqref{ttlikesol1} and \eqref{ttlikesol2} as expected. A simple example of this behaviour is \eqref{solsigma}.
\section{Higher dimensions}\label{sechigh}
Motivated by the remarkable properties of the massive gravity method in two dimensions, we will now try to apply our methodology in higher dimensions. During this section, a lot of traces of powers of the stress-tensor will appear, hence we slightly switch our language adopting a matrix notation, leaving the indices behind. To avoid any misunderstanding, we will treat \(\tensor{T}{^{\mu}_{\nu}}\) as a matrix with the first index representing rows and the second columns. Furthermore, the coefficient \(b_{0}\) will appear extensively in our calculations, and at times, in the denominator of expressions, therefore we will always assume that it does not approach zero. This fact is verified a posteriori.
\subsection{Deformations of order three in three dimensions}\label{sec3d}
Looking back to \eqref{2ddef} and \eqref{gsol}, we found solutions for the most general quadratic operator using a quadratic ansatz for \(G\). In a direct analogy, we now consider the most general cubic deformation i.e.
\begin{gather}
	\begin{aligned}
	\mathcal{O}=\int\dd[3]{x}f\hspace{0.05cm}\Bigg{\{}\la b_{0}\tr T^{3}+\la b_{1}\tr T^{2}\tr T+&\la b_{2}(\tr T)^{3}+\\+b_{3}\tr T^{2}+&b_{4}(\tr T)^{2}+\dfrac{b_{5}}{\la}\tr T+\dfrac{b_{6}}{\la^{2}}\Bigg{\}}\,, \label{3dopdef}
		\end{aligned}
\end{gather}
where the dependence on \(\la\) is fixed such that all \(b_{i}\) are arbitrary and dimensionless. The mass units of \(\la\) are \(M^{-3}\). The second ingredient is the ansatz for \(G\), which we pick to be the most generic cubic polynomial in \((y_{1}, y_{2}, y_{3})\) or
\begin{gather}
	\begin{aligned}
	G(\la,\mathfrak{g}_{i},y_{1},y_{2},y_{3})=c_{0}(\la,\mathfrak{g}_{i})+c_{1}(\la,\mathfrak{g}_{i})y_{1}+c_{2}(\la,\mathfrak{g}_{i})y_{1}^{2}+c_{3}(\la,\mathfrak{g}_{i})y_{2}+\\+c_{4}(\la,\mathfrak{g}_{i})y_{1}^{3}+c_{5}(\la,\mathfrak{g}_{i})y_{1}y_{2}+c_{6}(\la,\mathfrak{g}_{i})y_{3}\,.
\end{aligned}
\end{gather}
From there, the massive gravity method yields the following solution 
\begin{gather}
	\begin{gathered}
		\begin{aligned}
			c_{0}(\la,\mathfrak{g}_{i})&=-\dfrac{4b_{3}}{3 b_{0}\la}\,,\quad c_{1}(\la,\mathfrak{g}_{i})=\mathfrak{c_{1}}(\mathfrak{g}_{i})\la^{b}\,,\quad c_{2}(\la,\mathfrak{g}_{i})=0=c_{3}(\la,\mathfrak{g}_{i})\,,\quad \\ c_{4}(\la,\mathfrak{g}_{i})&=\mathfrak{c_{4}}(\mathfrak{g}_{i})-\dfrac{\mathfrak{c_{1}}(\mathfrak{g}_{i})^{3}b_{0}}{4+6b}\la^{2+3b}\,,\quad
			c_{5}(\la,\mathfrak{g}_{i})=-3c_{4}(\la,\mathfrak{g}_{i})\,,\quad \\
				c_{6}(\la,\mathfrak{g}_{i})&=2c_{4}(\la,\mathfrak{g}_{i})\,,\quad b\coloneqq2b_{5}+\dfrac{4b_{3}^{2}}{3b_{0}}\,. \label{csol3d}
		\end{aligned} 
	\end{gathered}
\end{gather}
where \(b_{1}, b_{2}, b_{4}\) and \(b_{6}\) are now constrained,
\begin{gather}
	b_{1}=-\dfrac{3b_{0}}{4}\,,\quad b_{2}=\dfrac{b_{0}}{8}\,,\quad b_{4}=-\dfrac{b_{3}}{2}\,,\quad b_{6}=4\dfrac{3b_{0}b_{3}+4b_{3}^{3}+9b_{0}b_{3}b_{5}}{9b_{0}^{2}}\,. \label{bis3d}
\end{gather}
Surprisingly, the ratio \(b_{4}/b_{3}\) of the quadratic terms in the deformation, is equal to \(-1/(3-1)\) in agreement with \cite{Taylor:2018xcy,Hartman:2018tkw}. We should also note that with this choice of \(b_{1}\) and \(b_{2}\), the cubic order terms of the operator are not equal to the determinant of the stress-tensor. From a dimensional standpoint, since \(G\) has mass units \(M^{3}\) the dimensions of \(\mathfrak{c_{1}}\) and \(\mathfrak{c_{4}}\) are uniquely determined
	\begin{gather}
		\qty[\mathfrak{c_{1}}(\mathfrak{g}_{i})]=M^{3+3b}\,, \qquad \qty[\mathfrak{c_{4}}(\mathfrak{g}_{i})]=M^{3}\,.     \label{diman3d}
\end{gather}
 In order to simplify things, let us consider the case where no other couplings are present, which translates to the condition \(b=-1\) and \(\mathfrak{c_{4}}=0\). This gives yet another constraint
\begin{gather}
	b_{5}=-\dfrac{1}{2}-\dfrac{2b_{3}^{2}}{3b_{0}}\,, \label{3dcon}
\end{gather}
thus the massive gravity action assumes a familiar form \eqref{soll2d}
\begin{gather}
		S_{\text{G}}=\int\dd[3]{x}\dfrac{f}{\la'}\qty{-4b_{3}+3b_{0}\mathfrak{c_{1}}y_{1}+(3b_{0})^{2}\mathfrak{c_{1}}^{\!\!\!3}\det Y_{1}}\,, \label{soll3d}
\end{gather} 
where \(\la'\coloneqq3\la b_{0}\). Once again, this theory can be fully determined by applying the initial condition \eqref{inco}, which fixes \(\mathfrak{c_{1}}\) exactly. Moreover, we observe that the last term in the equation above, completely decouples from the background as in the two-dimensional case. The full solution now reads
\begin{gather}
		S[\tensor{e}{_{\m}^{a}},\tensor{f}{_{\m}^{a}},\la,\mathfrak{g}_{i},\Phi]=\int\dd[3]{x}\dfrac{f}{\la'}\qty{3b_{0}\mathfrak{c_{1}}y_{1}-4b_{3}}+(3b_{0})^{2}\mathfrak{c_{1}}^{\!\!\!3}\int\dd[3]{x}\dfrac{e}{\la'}+S_{0}[\tensor{e}{_{\m}^{a}},\mathfrak{g}_{i},\Phi]\,, \label{fullsoll3d}
\end{gather} 
where one notices that, essentially, all the dynamics of the gravity action reduce to just \(y_{1}\), whereas the seed theory absorbs a constant term proportional to the inverse of \(\la'\). We must note that this is a simple rewriting of the terms and it does not change the flow equation or the deforming operator. A first quick calculation regarding the operator \eqref{3dopdef} is to find its trace-flow equation. For the massive gravity solution \eqref{soll3d} and using (\ref{wa}, \ref{eqmot}) we retrieve
\begin{gather}
	\begin{aligned}
		\dfrac{\bar{e}\tr\bar{\mathcall{S}}}{3}=\dfrac{f\tr\bar{T}}{3}+f\la\Bigg{\{}\la b_{0}\tr\bar{T}^{3}+\la b_{1}\tr\bar{T}^{2}\tr\bar{T}+&\la b_{2}(\tr\bar{T})^{3}+\\+b_{3}\tr\bar{T}^{2}+&b_{4}(\tr\bar{T})^{2}+\dfrac{b_{5}}{\la}\tr\bar{T}+\dfrac{b_{6}}{\la^{2}}\Bigg{\}}\,, \label{traceflow3d}
	\end{aligned}
\end{gather}
therefore if the seed theory does not contain any dimensionful couplings, the combination of \eqref{3dopdef} and \eqref{traceflow3d} leads to the following flow equation
\begin{gather}
	\dv{S}{\la}=-\dfrac{1}{3\la}\int\dd[3]{x}f\hspace{0.05cm}\tr\bar{T}=\dfrac{1}{\Delta_{\la}\la}\int\dd[3]{x}f\hspace{0.05cm}\tr\bar{T}\,,
\end{gather}
when we are on-shell for the auxiliary dreibein, where \(\Delta_{\la}\) is the mass dimension of \(\la\). This is exactly what one would expect for a theory with only one dimensionful coupling.

Following this analysis, we would like to explore if more choices exist for the \(b_{i}\) which are not predicted by the massive gravity method. In order to achieve this, we will consider a simple Nambu-Goto boson in \(d\)-dimensions described by the following action
\begin{gather}
	S=\int\dd[d]{x}\dfrac{f}{\xi\la}\qty{\rho-\sqrt{1+\xi \la \partial^{\mu}\phi\partial_{\mu}\phi}}\,, \label{NGactiond}
\end{gather} 
where \(\xi\) and \(\rho\) are dimensionless constants, and check if it is a solution of the deforming operator \eqref{3dopdef} for \(d=3\). Interestingly, we find that this model is indeed flowing with respect to this deformation for the following choices of \(b_{i}\)
\begin{gather}
	\begin{aligned}
		b_{0}&=\dfrac{\xi^{2}}{6}\,,\qquad b_{1}=-\dfrac{\xi^{2}}{4}\,,\qquad b_{2}=\dfrac{\xi^{2}}{12}\,,\qquad b_{3}=\dfrac{\rho\xi}{4}\,,\qquad \\ b_{4}&=-\dfrac{\rho\xi}{4}\,,\qquad b_{5}=\dfrac{1}{2}(\rho^{2}-1)\,,\qquad b_{6}=\dfrac{\rho}{2\xi}(1-\rho^{2})\,.\qquad
	\end{aligned}
\end{gather} 
This solution is not consistent with the constraints \eqref{bis3d} and \eqref{3dcon} therefore we conclude that more deformations of cubic order with the same structure as in \eqref{3dopdef} exist. 
\subsection{Deformations of order four in four dimensions}\label{sec4d}
Let us now repeat the same process in four dimensions, with the hope of extracting the general behaviour of the deformation in arbitrary dimensions. We begin from an ansatz of order four for the deformation
\begin{gather}
	\begin{aligned}
		\mathcal{O}=\int\dd[4]{x}f\hspace{0.05cm}\Bigg{\{}\la^{2}b_{0}\tr T^{4}+\la^{2}b_{1}\tr T^{3}\tr T+\la^{2} b_{2}(\tr T^{2})^{2}+\la^{2} b_{3}\tr T^{2}(\tr T)^{2}+&\\+\la^{2} b_{4}(\tr T)^{4}+\la b_{5}\tr T^{3}+\la b_{6}\tr T^{2}\tr T+\la b_{7}(\tr T)^{3}+&\\ +b_{8}\tr T^{2}+ b_{9}(\tr T)^{2}+\dfrac{b_{10}}{\la}\tr T+\dfrac{b_{11}}{\la^{2}}&\Bigg{\}}\,, \label{4dopdef}
	\end{aligned}
\end{gather}
where all \(b_{i}\) are unitless. The mass units of \(\la\) in this case are \(M^{-4}\) and \(M^{4}\) for \(G\). Looking for solutions of quartic form
\begin{gather}
	\begin{aligned}
		G(\la,\mathfrak{g}_{i},y_{1},y_{2},y_{3},y_{4})=c_{0}(\la,\mathfrak{g}_{i})+c_{1}(\la,\mathfrak{g}_{i})y_{1}+c_{2}(\la,\mathfrak{g}_{i})y_{1}^{2}+c_{3}(\la,\mathfrak{g}_{i})y_{2}+&\\+c_{4}(\la,\mathfrak{g}_{i})y_{1}^{3}+c_{5}(\la,\mathfrak{g}_{i})y_{1}y_{2}+c_{6}(\la,\mathfrak{g}_{i})y_{3}+c_{7}(\la,\mathfrak{g}_{i})y_{1}^{4}+&\\+c_{8}(\la,\mathfrak{g}_{i})y_{1}^{2}y_{2}+c_{9}(\la,\mathfrak{g}_{i})y_{1}y_{3}+c_{10}(\la,\mathfrak{g}_{i})y_{2}^{2}+c_{11}(\la,\mathfrak{g}_{i})y_{4}\,,
	\end{aligned}
\end{gather}
we find
\begin{gather}
	\begin{gathered}
		\begin{aligned}
			c_{0}(\la,\mathfrak{g}_{i})&=-\dfrac{9b_{5}}{4\la b_{0}}\,,\quad c_{1}(\la,\mathfrak{g}_{i})=\mathfrak{c_{1}}(\mathfrak{g}_{i})\la^{b}\,,\quad c_{j}(\la,\mathfrak{g}_{i})=0\,,\quad j=\{2,3,4,5,6\}\,, \\ c_{7}(\la,\mathfrak{g}_{i})&=\mathfrak{c_{7}}(\mathfrak{g}_{i})-\dfrac{\mathfrak{c_{1}}(\mathfrak{g}_{i})^{4}b_{0}}{6(3+4b)}\la^{3+4b}\,,\quad
			c_{8}(\la,\mathfrak{g}_{i})=-6c_{7}(\la,\mathfrak{g}_{i})=c_{11}(\la,\mathfrak{g}_{i})\,,\quad \\
		\end{aligned} \\
		c_{9}(\la,\mathfrak{g}_{i})=8c_{7}(\la,\mathfrak{g}_{i})\,, \quad c_{10}(\la,\mathfrak{g}_{i})=3c_{7}(\la,\mathfrak{g}_{i})\,, \quad
		b\coloneqq 3b_{10}+\dfrac{27b_{5}^{3}}{16b_{0}^{2}}\,, \label{csol4d}
	\end{gathered}
\end{gather}
where many of the \(b_{i}\) are constrained in the following way
\begin{gather}
	\begin{aligned}
		b_{1}=-\dfrac{8b_{0}}{9}\,,\quad b_{2}=-\dfrac{b_{0}}{2}\,,\quad b_{3}=\dfrac{5b_{0}}{9}\,,\quad b_{4}=-\dfrac{11b_{0}}{162}\,,\quad
		b_{6}=-\dfrac{b_{5}}{2}\,,\quad	\\
		b_{7}=\dfrac{b_{5}}{18}\,,\quad b_{8}=\dfrac{9b_{5}^{2}}{8b_{0}}\,,\quad 
		b_{9}=-\dfrac{3b_{5}^{2}}{8b_{0}}\,,\quad  
		b_{11}=9\dfrac{27b_{5}^{4}+16b_{0}^{2}b_{5}(1+4b_{10})}{64b_{0}^{3}}. \label{4dcoef}
	\end{aligned}
\end{gather}
From the constraints above, we notice that the ratio \(b_{9}/b_{8}\) is equal to \(-1/(4-1)\) confirming once more \cite{Taylor:2018xcy,Hartman:2018tkw}. Moreover, the choice above does not imply that the quartic order terms are proportional to the determinant. The units of \(\mathfrak{c_{1}}\) and \(\mathfrak{c_{7}}\) are given by
\begin{gather}
	\qty[\mathfrak{c_{1}}(\mathfrak{g}_{i})]=M^{4+4b}\,, \qquad \qty[\mathfrak{c_{7}}(\mathfrak{g}_{i})]=M^{4}\,.  \quad   \label{diman4d}
\end{gather}
Furthermore, assuming that \(\la\) is the only dimensionful coupling, \(b=-1\) and \(\mathfrak{c_{7}}=0\) just like the three-dimensional case. The extra constraint now reads
\begin{gather}
	b_{10}=-\dfrac{1}{3}-\dfrac{9b_{5}^{3}}{16b_{0}^{2}}\,, \label{4dcon}
\end{gather}
and the massive gravity action follows
\begin{gather}
		S_{\text{G}}=\int\dd[4]{x}\dfrac{f}{\la'}\qty{-9b_{5}+4b_{0}\mathfrak{c_{1}}y_{1}+(4b_{0})^{2}\mathfrak{c_{1}}^{\!\!\!4}\det Y_{1}}\,, \label{soll4d}
\end{gather}
where \(\la'\coloneqq4\la b_{0}\) and \(\mathfrak{c_{1}}\) is fixed by \eqref{inco}. 
The full solution becomes
\begin{gather}
	S[\tensor{e}{_{\m}^{a}},\tensor{f}{_{\m}^{a}},\la,\mathfrak{g}_{i},\Phi]=\int\dd[4]{x}\dfrac{f}{\la'}\qty{4b_{0}\mathfrak{c_{1}}y_{1}-9b_{5}}+(4b_{0})^{2}\mathfrak{c_{1}}^{\!\!\!4}\int\dd[4]{x}\dfrac{e}{\la'}+S_{0}[\tensor{e}{_{\m}^{a}},\mathfrak{g}_{i},\Phi]\,, \label{fullsoll4d}
\end{gather}
where the same behaviour as in three dimensions is noted. We proceed with the trace-flow equation, which assumes the following familiar form
\begin{gather}
	\begin{aligned}
		\dfrac{\bar{e}\tr\bar{\mathcall{S}}}{4}=\dfrac{f\tr\bar{T}}{4}+f\la\Bigg{\{}\la^{2}b_{0}\tr\bar{T}^{4}+\la^{2}b_{1}\tr\bar{T}^{3}\tr\bar{T}+\la^{2} b_{2}(\tr\bar{T}^{2})^{2}+\la^{2} b_{3}\tr\bar{T}^{2}(\tr \bar{T})^{2}+&\\+\la^{2} b_{4}(\tr\bar{T})^{4}+\la b_{5}\tr\bar{T}^{3}+\la b_{6}\tr\bar{T}^{2}\tr\bar{T}+\la b_{7}(\tr\bar{T})^{3}+&\\ +b_{8}\tr\bar{T}^{2}+ b_{9}(\tr \bar{T})^{2}+\dfrac{b_{10}}{\la}\tr\bar{T}+\dfrac{b_{11}}{\la^{2}}\Bigg{\}}\,,& \label{traceflow4d}
	\end{aligned}
\end{gather}
therefore when the auxiliary vierbein is eliminated and \(\la\) is the only dimensionful coupling, the four-dimensional flow equation is
\begin{gather}
	\dv{S}{\la}=-\dfrac{1}{4\la}\int\dd[4]{x}f\hspace{0.05cm}\tr\bar{T}=\dfrac{1}{\Delta_{\la}\la}\int\dd[4]{x}f\hspace{0.05cm}\tr\bar{T}\,,
\end{gather}
where yet again, \(\Delta_{\la}\) is the mass dimension of \(\la\) as expected.

 Lastly, let us check which are the solutions that \eqref{NGactiond} predicts at \(d=4\) given the operator \eqref{4dopdef}. We find that \(b_{0}, b_{1}\) and \(b_{2}\) are unconstrained, therefore we can freely set them to zero. This choice conveniently leads to \(b_{3}=0=b_{4}\), which removes the quartic order terms entirely from the operator. The surviving coefficients read
\begin{gather}
	\begin{aligned}
		b_{5}&=\dfrac{\xi^{2}}{8}\,,\qquad b_{6}=-\dfrac{\xi^{2}}{6}\,,\qquad b_{7}=\dfrac{\xi^{2}}{24}\,,\qquad b_{8}=\dfrac{7\rho\xi}{24}\,,\qquad \\ b_{9}&=-\dfrac{\rho\xi}{6}\,,\qquad b_{10}=\dfrac{3}{8}(\rho^{2}-1)\,,\qquad b_{11}=\dfrac{\rho}{2\xi}(1-\rho^{2})\,.\qquad
	\end{aligned}
\end{gather} 
Once again, the coefficients above do not match with \eqref{4dcoef} and \eqref{4dcon}, which suggests the existence of other operators with the same form besides \eqref{4dopdef}.

 The analysis of these two sections is quite tedious, but provides a crucial insight to the structure of these deformations which is necessary for the uplift in \(d\) dimensions. 
\subsection{Five, six, seven and \(d\) dimensions}
Contrary to the title of this section, we will begin the discussion directly in \(d\) dimensions, and then confirm everything in dimension five, six and seven. Let us first fix the mass units of \(\la\) to be \(M^{-d}\) in accordance with every section that precedes this one and revisit equations \eqref{diman3d} and \eqref{diman4d}. We assume that in \(d\) dimensions there does exist a similar set of two \(\mathfrak{c_{i}}\) which we will refer to as \(\mathfrak{c_{1}}\) and \(\mathfrak{c_{\star}}\) with dimensions
\begin{gather}
		\qty[\mathfrak{c_{1}}(\mathfrak{g}_{i})]=M^{d+db}\,, \qquad \qty[\mathfrak{c_{\star}}(\mathfrak{g}_{i})]=M^{d}\,.   \label{dimandd}
\end{gather}
where \(b\) is a complicated function of \(b_{i}\) and \(d\). Looking at the dimensionality of \(\mathfrak{c_{\star}}\), we understand that there are only four ways it can be part of the massive gravity action. The first possibility is the simplest one, it just enters only by coupling to \(y_{i}\). The other options are: an appropriate function of \((\la,\mathfrak{c_{\star}})\) or \((\mathfrak{c_{1}},\mathfrak{c_{\star}})\) or \((\la,\mathfrak{c_{1}},\mathfrak{c_{\star}})\), all coupled to \(y_{i}\). A natural assumption is that the deformation should be the same for any seed theory carrying the couplings \(\mathfrak{g}_{i}\) including theories where \(\mathfrak{g}_{i}\rightarrow 0\). The only choice that is consistent with that limit is the first one, and it also constrains the way it couples to \(y_{i}\). We deduce that \(\mathfrak{c_{\star}}\) enters the \(d\)-dimensional massive gravity solution in the following way
\begin{gather}
	\int\dd[d]{x}f\hspace{0.05cm}d!\mathfrak{c_{\star}}(\mathfrak{g}_{i})\det Y_{1}=\int\dd[d]{x}e\hspace{0.05cm}d!\mathfrak{c_{\star}}(\mathfrak{g}_{i})\,,
\end{gather} 
which agrees with our previous treatments in two, three and four dimensions. From there it follows that \(\mathfrak{c_{\star}}\) essentially represents the seed theory \(S_{0}\). To make this argument more transparent, recall that all \(\mathfrak{c}_{i}\) were introduced as integration constants with respect to \(\la\) which means that they can, in principle, depend on the local coordinates. We now understand better the form of the generic solution \eqref{sol1}, which is consistent with the argument above. Since we are interested only in the massive gravity part of the solution we can safely turn off \(\mathfrak{c_{\star}}\) for the rest of this section.

The situation is not that simple for \(\mathfrak{c_{1}}\). Using the arguments above, \(\mathfrak{c_{1}}\) cannot appear on its own, we expect some coupling to \(\la\) through a unit-appropriate function. Nevertheless, we can however say more when the initial condition \eqref{inco} is imposed. In that case \(\mathfrak{c_{1}}\) is a dimensionless constant and \(b=-1\). Revisiting \eqref{soll3d} and \eqref{soll4d}, the last thing that we need to fix is the appearance of the constant term. We find that in the \(d\)-dimensional case, if one extends the deformation structure in precisely the same way, this constant is proportional to \(b_{p(d)}\) where \(p(d)\) is equal to the integer partitions of \(d\), and is controlling the term \(\tr T^{d-1}\). Combining all of the above, we propose the following solution in \(d\)-dimensions
\begin{gather}
	S_{\text{G}}=\int\dd[d]{x}\dfrac{f}{\la'}\qty{-(d-1)^{2}b_{p(d)}+db_{0}\mathfrak{c_{1}}y_{1}+(db_{0})^{2}\mathfrak{c_{1}}^{\!\!\!d}\det Y_{1}}\,, \qquad d\geq3\,, \label{solldd}
\end{gather}
where \(\la'\coloneqq d\la b_{0}\). Using this solution, the ratio \(r\) of the quadratic coefficients of the deformation is fixed, as well as the linear and constant coefficients. In more detail, let \(\be_{l}\) be the coefficient of \(\tr T\) and \(\be_{c}\) of the constant term of the deformation. Then, crucially, \(r=-1/(d-1)\) in agreement with \cite{Taylor:2018xcy,Hartman:2018tkw}, and
\begin{gather}
	 \be_{l}=\dfrac{1}{1-d}\qty(1+\dfrac{[(d-1)b_{p(d)}]^{d-1}}{[db_{0}]^{d-2}})\,, \qquad \be_{c}=\dfrac{(1-d)^{2}b_{p(d)}}{db_{0}} \be_{l}\,, \qquad d\geq3\,. \label{linconbe}
\end{gather}
The higher order coefficients of the deformation are unique functions of \(b_{0}\) and \(b_{p(d)}\), and can be evaluated by specifying the dimension, but a closed form expression is eluding us. It is rather odd that the massive gravity solution can be derived easier than the deformation itself, which is what we are interested in at the end of the day. Let us stop for a moment to comment on the last two equations. The main observation is the condition \(d\geq3\) which is not obvious, since we are free to set \(d=2\). This choice provides only a very specific solution in two dimensions and this is why we exclude it. In more detail setting \(d=2\) we observe that interestingly, \eqref{linconbe} yields \eqref{con2} with \(\be_{c}\leftrightarrow b_{3}\) as expected, but it constraints \(b_{2}\) specifically at the value \(b_{2}=-1/2\), which is the limit that removes the overall constant from the spectrum \eqref{solsprecmain}. Looking only at \eqref{solldd} for \(\mathfrak{c_{1}}=-1/(2b_{0})\), and comparing with \eqref{soll2d} we get the same constraint. Considering these facts, we conclude that \eqref{solldd} does not cover the entire span of solutions in two dimensions, hence the condition \(d\geq3\). There is however another interpretation in which the massive gravity solution \eqref{solldd} is more fundamental, meaning that in two dimensions, \(b_{2}\) must be equal to \(-1/2\).  In that case, the solution is valid for \(d\geq2\).  
 
In order to better understand the structure of the coefficients of the higher order terms in the deformation, we leave \(d=2\) behind us and assume that \(d\geq3\). Moreover, in order to highlight the importance of the quadratic terms, and to avoid using the awkward coefficient \(b_{p(d)}\), we define \(\be\) as
\begin{gather}
	b_{p(d)}\eqqcolon\dfrac{(2\be)^{\frac{1}{d-2}}(db_{0})^{\frac{d-3}{d-2}}}{d-1}\,, \label{bchoi}
\end{gather}
with \eqref{solldd} assuming the following form
\begin{gather}
	S_{\text{G}}=\int\dd[d]{x}\dfrac{f}{\la'}\qty{-(2\beta)^{\frac{1}{d-2}}(d-1)(d b_{0})^{\frac{d-3}{d-2}}+d b_{0}\mathfrak{c_{1}}y_{1}+(d b_{0})^{2}\mathfrak{c_{1}}^{\!\!\!d}\det Y_{1}}\,,  \label{solldbb}
\end{gather}
and the coefficients \(\be_{l}, \be_{c}\) given by \eqref{linconbe}, become
\begin{gather}
	\be_{l}=\dfrac{1+(2\be)^{\frac{1}{d-2}}(db_{0})^{\frac{d-3}{d-2}}}{1-d}\,, \qquad \be_{c}=-(d-1)\be_{l}\dfrac{1+(d-1)\be_{l}}{2\be}\,. \label{linconbffe}
\end{gather}
The first thing that we notice is that \(\be_{c}\) has a similar structure as in \eqref{con2} with the only difference being that \(\be_{l}\) is multiplied by \((d-1)\). Furthermore, the choice \eqref{bchoi} brings the deforming operator in the following form
\begin{gather}
		\mathcal{O}=\int\dd[d]{x}f\hspace{0.05cm}\Bigg{\{}\text{higher order terms}+\beta\qty(\tr T^{2}-\dfrac{1}{d-1}(\tr T)^{2})+\dfrac{\beta_{l}}{\la}\tr T +\dfrac{\beta_{c}}{\la^{2}}\Bigg{\}}\,, \label{dddopdef}
\end{gather}
  where the quadratic terms are in agreement with \cite{Taylor:2018xcy,Hartman:2018tkw}. As we mentioned earlier, the higher order terms are uniquely specified by inserting the solution and the operator above in the flow equation. For all practical purposes, and in order to verify that the massive gravity action \eqref{solldbb} is indeed valid, the deforming operators in five, six and seven dimensions as well as their higher order coefficients are given in appendix \ref{bees}. Another property of the operator above emerges if one compares (\ref{trflow}, \ref{traceflow3d}) and \eqref{traceflow4d}. It should not come as a surprise that the trace-flow equation now reads
 \begin{gather}
 		\dfrac{\bar{e}\tr\bar{\mathcall{S}}}{d}=\dfrac{f\tr\bar{T}}{d}+f\la\Bigg{\{}\text{\textoverline{higher order terms}}+\beta\qty(\tr\bar{T}^{2}-\dfrac{1}{d-1}(\tr\bar{T})^{2})+\dfrac{\beta_{l}}{\la}\tr\bar{T} +\dfrac{\beta_{c}}{\la^{2}}\Bigg{\}}\,, \label{traceflowdd}
 \end{gather}
 where the auxiliary vielbein is on-shell. A quick check in higher dimensions verifies that this equation holds. Then, turning off all couplings except \(\la\) returns the following flow equation
 \begin{gather}
 	\dv{S}{\la}=-\dfrac{1}{d\la}\int\dd[d]{x}f\hspace{0.05cm}\tr\bar{T}=\dfrac{1}{\Delta_{\la}\la}\int\dd[d]{x}f\hspace{0.05cm}\tr\bar{T}\,, \label{oncouple}
 \end{gather} 
 with  \(\Delta_{\la}\) being the mass dimension of \(\la\). 
 
 Let us attempt to investigate which are the potential necessary conditions for the stress-tensor, such that all higher order terms vanish. In \cite{Taylor:2018xcy,Hartman:2018tkw}, the constraint\footnote{Here we slightly abuse our notation since \(G\) is taken to be the massive gravity density.} \(\tensor{G}{^{0}_{i}}=0=\kappa^{2}\tensor{T}{^{0}_{i}}\) is extensively used therefore, this would be our starting point. This implies that \(\tensor{T}{^{i}_{0}}=0\) also,\footnote{We notice that this procedure shares some common ground with the dimensional reduction followed in subsection \ref{secdimred}.} therefore the \(d\)-dimensional stress-tensor assumes the following block diagonal form
 \begin{gather}
 	\tensor[^{(d)}]{T}{^{\mu}_{\nu}}=\tensor[^{(d)}]{T}{^{0}_{0}}\oplus\tensor[^{(d-1)}]{T}{^{\al}_{\be}}.
 \end{gather}
Then, we find that setting
\begin{gather}
	\tensor[^{(d)}]{T}{^{0}_{0}}=\dfrac{1}{d-2}\tensor[^{(d-1)}]{T}{^{\al}_{\al}}\,, \qquad \tensor[^{(d)}]{\be}{}=\dfrac{1}{2}(d-1)^{d-2}d^{3-d}\tensor[^{(d)}]{b}{_{0}}\,, \label{dimredcond}
\end{gather}
the \(d\)-dimensional operator reduces to the \((d-1)\)-dimensional one, at least for all the cases that we have studied, excluding \(d=1\). As an example, let us consider the linear term of the deformation but now in \(d\)-dimensions. Using the condition above, the trace of the stress-tensor and the coefficient \(\be_{l}\) satisfy
\begin{gather}
\tensor[^{(d)}]{T}{^{\mu}_{\mu}}=\dfrac{d-1}{d-2}\tensor[^{(d-1)}]{T}{^{\al}_{\al}}, \qquad	\tensor[^{(d)}]{\be}{_{l}}=\dfrac{d-2}{d-1}\tensor[^{(d-1)}]{\be}{_{l}}\,, \label{dimredstress}
\end{gather}
thus it is clear that their product preserves its form in one dimension less i.e.
\begin{gather}
	\tensor[^{(d)}]{\be}{_{l}}\tensor[^{(d)}]{T}{^{\mu}_{\mu}}=\tensor[^{(d-1)}]{\be}{_{l}}\tensor[^{(d-1)}]{T}{^{\al}_{\al}}.
\end{gather}
This is precisely the behaviour we observe. Assuming now that \eqref{dimredcond} holds for every \(d\geq2\), all higher order terms vanish identically, leaving us only with the quadratic deformation.  We should also note that in the last iteration of this process we recover the two-dimensional solution \eqref{opp2d}, with \(b_{2}=-1-4b_{0}/3\). Moreover, the recursive relation \eqref{dimredstress} for the stress-tensor can be solved yielding
\begin{gather}
	\tensor[^{(d)}]{T}{^{\mu}_{\mu}}=(d-1)\tensor[^{(2)}]{T}{^{\al}_{\al}}\,, \label{stress2sol}
\end{gather}
where the initial condition is located at \(d=2\) as we discussed. This equation suggests that a hidden two-dimensional stress tensor is the source of the deformation. A more heuristic argument for these choices will be presented during the discussion in section \ref{disc}.   
\subsection{Root deformations}
For this final subsection, we will consider the following root-\(T\overline{T}\) operator
\begin{gather}
	\mathcal{R}=\int\dd[d]{x}f\hspace{0.05cm}b_{\text{r}}\sqrt{\tensor{T}{^\mu_\nu}\tensor{T}{^\nu_\mu}-\frac{1}{d}(\tensor{T}{^\mu_\mu})^{2}}\,, \label{ttlikemaind} 
\end{gather}
in direct analogy with \eqref{ttlikemain}, and use the massive gravity method in order to understand its behaviour. Our guiding principle is the ModMax theory in four dimensions \cite{Bandos:2020jsw,Floss:2023nod}, which is part of the flow of the operator above at \(d=4\) \cite{Babaei-Aghbolagh:2022uij,Ferko:2022iru,Babaei-Aghbolagh:2022leo}. The Lagrangian density for this model is given by
\begin{gather}
	\mathcal{L}_{\text{ModMax}}=\mathcal{L}_{\text{Max}}\cosh(b_{\text{r}}\m)+\sqrt{\mathcal{L}_{\text{Max}}^{2}+\mathcal{L}_{\text{p}}^{2}}\sinh(b_{\text{r}}\m)\,, \label{modmax}
\end{gather}
where \(\mathcal{L}_{\text{Max}}\coloneqq-\tr F^{2}/4\) and \(\mathcal{L}_{p}\coloneqq-\tr(F\star F)/4\), with \(\tensor{F}{_{\mu}_{\nu}}\) being the usual electromagnetic tensor and \(\tensor{(\star F)}{_{\mu}_{\nu}}\) its Hodge-dual. We observe that only the two aforementioned variables are present, which indicates that out of the set of variables \(y_{i}\) \eqref{genop}, with \(1\leq i\leq d\), for our massive gravity formalism, there must be choice such that only two survive. Taking into account that  \(\tensor{F}{_{\mu}_{\nu}}\) is antisymmetric, we are naturally lead to an antisymmetric choice for \(Y_{1}\) which in turn restricts the massive gravity variables to \(y_{2i}\) with \(1\leq i\leq \lfloor d/2\rfloor\). The two-variable choice and the floor function, restrict the number of dimensions to \(d=4\) or \(d=5\). In both cases, the surviving variables are \(y_{2}\) and \(y_{4}\), yet their flow equations for \eqref{ttlikemaind} are drastically different. Specifically at \(d=5\), we were not able to find any solutions. Settling for four dimensions, the flow equation assumes the following familiar form
\begin{gather}
	\partial_{\mu}G=b_{\text{r}}\sqrt{4y_{4}-y_{2}^{2}}(\partial_{y_{2}}G+y_{2}\partial_{y_{4}}G)\,, \label{ttlikepdemain4}
\end{gather} 
where \(b_{\text{r}}\) has absorbed a \(\pm\) factor just like in the two-dimensional case. The equation above can be solved exactly
\begin{gather}
	G(\mu,\mathfrak{g}_{i},y_{2},y_{4})=\mathcall{G}\qty[\mathfrak{g}_{i},y_{2}^{2}-2y_{4},\mu b_{\text{r}}+\arctanh\qty(y_{2}^{-1}\sqrt{4y_{4}-y_{2}^{2}})]\,, \label{ttlikemainsol4}
\end{gather}
for any function \(\mathcall{G}\), which is strikingly similar to \eqref{ttlikemainsol}. We again observe that the deformation parameter decouples from the massive gravity variables, therefore the same arguments as in subsection \ref{secroot} apply. This should also indicate the presence of ``seed'' massive gravity, which in turn means that the uplift \(y_{2}\rightarrow y_{2,\mu}\) and \(y_{4}\rightarrow y_{4,\mu}\) should exist. Remarkably, all these statements are true since the solution can be brought in the following form
\begin{gather}
	G(\mu,\mathfrak{g}_{i},y_{2},y_{4})=\mathcall{G}\qty(\mathfrak{g}_{i},y_{2,\m},y_{4,\m})\,, \label{ttlikesol1main4}
\end{gather}
where
\begin{gather}
	\begin{aligned}
		&y_{2,\m}=y_{2}\cosh(b_{\text{r}}\m)+\sqrt{4y_{4}-y_{2}^{2}}\sinh(b_{\text{r}}\m)\,,\\&y_{4,\m}=y_{4}\cosh(2b_{\text{r}}\m)+\dfrac{y_{2}}{2}\sqrt{4y_{4}-y_{2}^{2}}\sinh(2b_{\text{r}}\m)\,.
	\end{aligned} \label{ttlikesol2main4}
\end{gather} 
Moreover, the second argument in \eqref{ttlikemainsol4} indicates that \(2y_{4,\m}-y_{2,\m}^{2}=2y_{4}-y_{2}^{2}\), which is verified easily using the equations above. The similarities do not stop there, we notice that \eqref{ttlikesol2main4} remains invariant under \(y_{2}\leftrightarrow y_{2,\mu}\) and \(y_{4}\leftrightarrow y_{4,\mu}\), therefore by defining the vierbein \(\tensor{\varepsilon}{_{\mu}^{a}}\) such that
\begin{gather}
	y_{2}^{\star}\coloneqq\tensor{f}{^{\nu}_{b}}\tensor{\varepsilon}{_{\la}^{b}}\tensor{f}{^{\la}_{c}}\tensor{\varepsilon}{_{\nu}^{c}}=y_{2,\mu}\,,\qquad y_{4}^{\star}\coloneqq\tensor{f}{^{\nu}_{b}}\tensor{\varepsilon}{_{\la}^{b}}\tensor{f}{^{\la}_{c}}\tensor{\varepsilon}{_{\rho}^{c}}\tensor{f}{^{\rho}_{d}}\tensor{\varepsilon}{_{\sigma}^{d}}\tensor{f}{^{\sigma}_{h}}\tensor{\varepsilon}{_{\nu}^{h}}=y_{4,\m} \,, \label{defroot4}
\end{gather}
one should be able to derive expressions that are equivalent to \eqref{root2dsolfin1} and \eqref{root2dsolfin} in four dimensions. We cannot help but notice that there does exist a set of equations given by  
\begin{gather}
	\begin{aligned}
		&\chi_{n,\m}=\chi_{n}\cosh(b_{\text{r}}\m)+\sqrt{2n\chi_{m}-\chi_{n}^{2}}\sinh(b_{\text{r}}\m)\,,\\&\chi_{m,\m}=\chi_{m}\cosh(2b_{\text{r}}\m)+\dfrac{\chi_{n}}{n}\sqrt{2n\chi_{m}-\chi_{n}^{2}}\sinh(2b_{\text{r}}\m)\,,
	\end{aligned} \label{ttlikesol2maind}
\end{gather} 
such that  \(n\chi_{m,\m}-\chi_{n,\m}^{2}=n\chi_{m}-\chi_{n}^{2}\) and remains invariant under the change \(\chi_{n}\leftrightarrow \chi_{n,\mu}\) and \(\chi_{m}\leftrightarrow \chi_{m,\mu}\). It would be interesting to explore if this generalization can be useful.

Let us now close this section by making the connection to the matter seed theory in four dimensions. If this type of deformation is universal, one would expect that upon correct identification of the generic scalars \(x_{2}\) and \(x_{4}\) in the seed theory \eqref{eq7}, the deformation should be given by the set \eqref{ttlikesol2main4}. Surprisingly, identifying \(x_{2}\leftrightarrow-\tr F^{2}/4\) and \(x_{4}\leftrightarrow\tr F^{4}/16\) returns \eqref{modmax}.  It is then clear that the antisymmetry of \(\tensor{F}{_{\mu}_{\nu}}\) implies the antisymmetry of \(Y_{1}\) and vice versa. 
\section{Summary and discussion} \label{disc}
In this work, we extended the massive gravity method in arbitrary dimensions. We began with a detailed explanation of the methodology for general \(d\). Briefly, we introduced the background and auxiliary vielbeins and a simple yet powerful reformulation of the deformed action in a massive gravity part constructed by the two kinds of vielbeins and a seed theory part, which depends exclusively on matter and the auxiliary vielbein. Once the deformation and the massive gravity action are specified, the auxiliary vielbein is integrated out, either classically or quantum mechanically. This approach comes in direct analogy to the well-studied two-dimensional case \cite{Tolley:2019nmm,Mazenc:2019cfg}.

  In terms of direct applications, setting our starting point in two dimensions, we derived the most general quadratic massive gravity solution using the most general ansatz of the same order as the deforming operator. We observed that the nature of the solution constrained the coefficients of the quadratic terms of the deforming operator to be exactly equal to \(T\overline{T}\), with the coefficient in front of the linear term being unaffected by that choice, something which was also true for the constant term. Then, we showed that \(T\overline{T}\) and \(T\overline{T}+\Lambda_{2}\) naturally drop out at certain limits of this deformation, indicating its generality. The next stop of this journey was the spectrum analysis of this generalized operator when no other dimensionful couplings were present. In this limit the operator is given by \eqref{opp2d}. The spectrum \eqref{solsprecmain} was found to be compatible with a string theory spectrum enriched by a \(B\)-field contribution under certain identifications. From there, the source of this extra term was found to essentially be the linear and constant terms that are present in our operator and can be turned off for the choice \(b_{2}=-1/2\). Notably, for this choice the limit \(\la\rightarrow 0\) is no longer well-defined.
  
   Another interesting property of \eqref{opp2d} was unveiled in subsection \ref{correl}. In short, we found that the coefficient in front of the linear term, can be uplifted to a source term for the deformed theory, allowing one to evaluate \(T\overline{T}\)-deformed correlators of undeformed operators either directly from the deformed partition function \eqref{corr1} or via the undeformed one \eqref{corr2}. It would be instructive to compare our results with other approaches \cite{Kraus:2018xrn,Cardy:2019qao,Cui:2023jrb,Aharony:2023dod,Barel:2024dgv}. Motivated by the uplift of the linear coefficient, we dimensionally reduced the operator using the generalized trace-flow equation \eqref{trflow} and matched it with the operator obtained by the dimensional reduction of a dilaton gravity action with an arbitrary potential \cite{Gross:2019ach} under certain non-trivial assumptions. The precise matching conditions are given by \eqref{matchhh}. Following this, the one dimensional massive gravity action was evaluated using the aforementioned methodology and a system of bosons and fermions was deformed yielding results consistent with the literature \cite{Gross:2019ach}. Returning to two dimensions, a similar classical system was deformed \cite{Cavaglia:2016oda,Kraus:2018xrn,Bonelli:2018kik,Frolov:2019nrr,Coleman:2019dvf,Caputa:2020lpa} yielding the more transparent expression \eqref{sol2dlin}. The first point we would like to stress further is that, in the process of deforming this type of seed actions we noticed that the intrinsic properties of matter do not enter the analysis, therefore one should only consider the specific way fields couple to the background. Furthermore, we showed that the deformation of the aforementioned model essentially allows for the deformation of higher-derivative theories. Finally, we studied the class of root-\(T\overline{T}\) deformations, finding that its structure does not admit a specific massive gravity solution, but enhances the solutions of other deformations \eqref{root2dsol}. After some manipulations, one can ``push'' this deformation on to the seed theory \eqref{rename2droot}, verifying that a massive gravity description is not necessarily needed.
  
  In higher dimensions, we began by deriving the operator in three and four dimensions, and used the solutions as our guiding principle. Then, an exact massive gravity prescription was found in \(d\) dimensions which reads
  \begin{gather}
  	S_{\text{G}}=\int\dd[d]{x}\dfrac{f}{\la'}\qty{-(2\beta)^{\frac{1}{d-2}}(d-1)(d b_{0})^{\frac{d-3}{d-2}}+d b_{0}\mathfrak{c_{1}}y_{1}+(d b_{0})^{2}\mathfrak{c_{1}}^{\!\!\!d}\det Y_{1}}\,, 
  \end{gather}
   where \(Y_{1}\) is given by \eqref{genop}, followed by the corresponding operator 
   \begin{gather}
   	\mathcal{O}=\int\dd[d]{x}f\hspace{0.05cm}\Bigg{\{}\text{higher order terms}+\beta\qty(\tr T^{2}-\dfrac{1}{d-1}(\tr T)^{2})+\dfrac{\beta_{l}}{\la}\tr T +\dfrac{\beta_{c}}{\la^{2}}\Bigg{\}}\,. 
   \end{gather}
     The higher order terms proved to be too cumbersome to determine for general \(d\), yet its quadratic structure precisely agrees with previous results \cite{Taylor:2018xcy, Hartman:2018tkw}. Moreover, for \(d=2\), we recover the operator \eqref{opp2d} with \(b_{2}=-1/2\), which is a choice that we found to be interesting. As a check, we derived the full solution up to seven dimensions. Following this, the generalized trace-flow equation \eqref{traceflowdd} is given, and a simple calculation \eqref{oncouple} verifies its validity. In an effort to reduce the order of the deformation, we found that the choice \(\tensor{T}{^{0}_{i}}=0=\tensor{T}{^{i}_{0}}\) along with
    \begin{gather}
    	\tensor[^{(d)}]{T}{^{0}_{0}}=\dfrac{1}{d-2}\tensor[^{(d-1)}]{T}{^{\al}_{\al}}\,, \qquad \label{disceq}
    \end{gather}
     for the operator in \(d\) dimensions, results in exactly the same operator in \((d-1)\) dimensions. Assuming that this is true for every \(d\geq3\), all greater than quadratic order terms vanish and the deformation matches \eqref{opp2d}. Then, the trace of the deformed stress-tensor in \(d\) dimensions is related to the two-dimensional one as indicated by \eqref{stress2sol}. In order to interpret this behaviour, consider a theory living in two dimensions along with its stress-tensor, which means that the  equation above gives \(\tensor[^{(3)}]{T}{^{0}_{0}}=\tensor[^{(2)}]{T}{^{\al}_{\al}}\). This essentially indicates that the third dimension is defined by the evolution of the theory in a direction normal to the two-dimensional manifold that it is living on. From the third-dimensional perspective, the equivalent statement is that the manifold is constructed by two-dimensional slices. Yet, \eqref{disceq} suggests that this is true for any \(d\), which means that one  has to iterate this process \((d-2)\) times in \(d\) dimensions. At each iteration, the energy content of the \((d-1)\)-dimensional theory is propagated in a direction normal to the manifold, effectively introducing the extra dimension. This is equivalent with the \((d-1)\)-dimensional slicing of a \(d\)-dimensional manifold. In addition, we notice that the coefficient in front of \(\tensor[^{(d-1)}]{T}{^{\al}_{\al}}\) in \eqref{disceq} demonstrates in which manner the two-dimensional energy content is distributed for general \(d\). If the arguments above hold, we conclude that the deformation in higher dimensions, is essentially given by the usual two-dimensional one, where the extra dimensions are generated by evolving the two-dimensional theory in \((d-2)\) normal directions in the way that we described. The last direction of research in this paper was the study of root-\(T\overline{T}\) deformations from the massive gravity perspective. We discovered that the only case that shares some key features with root-\(T\overline{T}\) in two dimensions is located at \(d=4\) in agreement with previous approaches \cite{Ferko:2022iru}.
  
   Let us briefly stress once more the importance of equations \eqref{dimandd}. During the early stages of this work, we assumed that the solution of the flow equation \eqref{floweq} assumes the form \(S_{\text{G}}+S_{0}\), where the \(\la\rightarrow 0\) limit returns only \(S_{0}\) evaluated on the background \(\tensor{f}{_{\m}^{a}}\). Notice that these are two conditions, essentially given by \eqref{inco}, one of which exclusively for \(S_{\text{G}}\). Since the massive gravity density satisfies the \((d+1)\)-dimensional partial differential equation \eqref{pde} and there is only one initial condition, one is restricted to the set of solutions that are fully determined by it. This is precisely the content of \eqref{dimandd}, with the constant \(\mathfrak{c_{1}}\) representing the initial condition for \(S_{\text{G}}\) and \(\mathfrak{c_{\star}}\) for \(S_{0}\). 
   
   Finally, we would like to make some concluding remarks about a potential generalization of our method. For the entirety of this work, we worked with operators that depended exclusively on conserved currents of translational symmetry i.e. the stress-tensor. This choice indicates that the background and auxiliary vielbeins must be used in the formulation of \(S_{\text{G}}\), since the stress tensor is generated by the variation of the action with respect to them. If some other conserved current, say \(J\), is used as part of the deformation, the auxiliary gauge field that generates \(J\) should be introduced in \(S_{\text{G}}\), where G now stands for gauge field. In two dimensions this is already true \cite{Tolley:2019nmm}, even in the non-relativistic limit \cite{Hansen:2020hrs}, and there is no obvious reason why this approach would not work in higher dimensions.   
\section*{Acknowledgments}
I would like to thank Davide Astesiano, Friðrik Freyr Gautason, Diego Hidalgo, Rahul Poddar, Domenico Seminara, Watse Sybesma, L\'arus Thorlacius, \'Ad\'am Tim\'ar, for interesting conversations, and especially Valentina Giangreco M. Puletti and Alexia Nix for extensive discussions and comments on the manuscript.  This work was supported by the
Icelandic Research Fund, grant 228952-052.
\newpage
\appendix
\section{Conventions} \label{conv}
Most of our notation agrees with \cite{Freedman:2012zz}. Everything in this section lives in \(d\) dimensions and we will always use the \((-,+,\dots,+)\) signature for the metric. Let indices \(M_{i}\) be either curved \((M_{i}\rightarrow\m_{i})\) or flat \((M_{i}\rightarrow m_{i})\). The (anti)symmetrization of a tensor \(A\) carrying \(n\) indices is defined as
\begin{gather}
\tensor{A}{_{[M_{1}}_\cdots_{M_{n}]}}\coloneqq\dfrac{1}{n!}\sum_{\s\in\mathfrak{S}_{n}}\sgn(\s)\tensor{A}{_{M_{\s(1)}}_\cdots_{M_{\s(n)}}}\,, \quad	\tensor{A}{_{(M_{1}}_\cdots_{M_{n})}}\coloneqq\dfrac{1}{n!}\sum_{\s\in\mathfrak{S}_{n}}\tensor{A}{_{M_{\s(1)}}_\cdots_{M_{\s(n)}}}\,,
\end{gather}
where \(\mathfrak{S}_{n}\) is the symmetric group of order \(n\). The Levi-Civita symbol \(\tensor{\e}{_{M_{1}}_{\cdots}_{M_{d}}}\) (or \(\tensor{\e}{^{M_{1}}^{\cdots}^{M_{d}}}\)) is exclusively used with \(\tensor{\e}{_0_{\cdots}_{d-1}}=\tensor{\e}{^0^{\cdots}^{d-1}}=1\). We also make use of the following identity
\begin{gather}
	\tensor{\e}{_{M_{1}}_{\cdots}_{M_{n}}_{N_{1}}_{\cdots}_{N_{d-n}}}\tensor{\e}{^{M_{1}}^{\cdots}^{M_{n}}^{R_{1}}^{\cdots}^{R_{d-n}}}=n!(d-n)!\tensor*{\de}{_{[N_{1}}^{R_{1}}}\cdots\tensor*{\de}{_{N_{d-n}]}^{R_{d-n}}}\,. \label{levci}
\end{gather}
Please note that the symbol with the indices up is independent from the one with the indices down. In order to relate the two one needs to define the Levi-Civita tensor as follows
\begin{gather}
	\tensor{\bar{\epsilon}}{_{M_{1}}_{\cdots}_{M_{d}}}\coloneqq\sqrt{\abs{\det[\tensor{g}{_{M}_{N}}]}}\tensor{\e}{_{M_{1}}_{\cdots}_{M_{d}}}\,,\quad	\tensor{\bar{\epsilon}}{^{M_{1}}^{\cdots}^{M_{d}}}\coloneqq\dfrac{\sgn(\det[\tensor{g}{_{M}_{N}}])}{\sqrt{\abs{\det[\tensor{g}{_{M}_{N}}]}}}\tensor{\e}{^{M_{1}}^{\cdots}^{M_{d}}}\,.
\end{gather}
The Clifford algebra is given by \(\{\tensor{\g}{^{a}},\tensor{\g}{^{b}}\}=2\tensor{\eta}{^a^b}\mathds{1}_{d}\), where \(\mathds{1}_{d}\) is the \(d\)-dimensional identity matrix. The \(\tensor{\g}{^{a}}\) matrices also satisfy \((\tensor{\g}{^{a}})^{\dagger}=\tensor{\g}{^{0}}\tensor{\g}{^{a}}\tensor{\g}{^{0}}\). Moreover \(\tensor{\g}{^{a_{1}}^\cdots^{a_{n}}}\coloneqq\tensor{\g}{^{[a_{1}}}\cdots\tensor{\g}{^{a_{n}]}}\). The Dirac adjoint is defined as \(\overline{\psi}\coloneqq\psi^{\dagger}i\g^{0}\) and the covariant derivatives on spinors as \(\vec{\nabla}_{\mu}\psi\coloneqq(\vec{\partial}_{\mu}+\frac{1}{4}\tensor{\om}{_{\mu}^{a}^{b}}\tensor{\g}{_{a}_{b}})\psi\) and \(\overline{\psi}\cev{\nabla}_{\mu}\coloneqq\overline{\psi}(\cev{\partial}_{\mu}-\frac{1}{4}\tensor{\om}{_{\mu}^{a}^{b}}\tensor{\g}{_{a}_{b}})\), where \(\tensor{\om}{_{\mu}^{a}^{b}}\) is the spin connection. Finally, \(A\vecev{\nabla}_{\mu}B\coloneqq A\vec{\nabla}_{\mu}B-A\cev{\nabla}_{\mu}B\).
\section{Characteristic and elementary symmetric polynomial(s)} \label{aapA}
The characteristic polynomial for any \(d\times d\) matrix \(A\) is given by 
\begin{gather}
	\det(\rho\mathds{1}_{d}-A)=\sum_{n=0}^{d}c_{n}\rho^{d-n}\,, \qquad c_{n}=(-1)^{n}e_{n}(\rho_{1},\rho_{2},\dots,\rho_{d})\,, \label{eqa1}
\end{gather}
where \(\mathds{1}_{d}\) is the \(d\)-dimensional identity matrix, \(\rho_{k}\) is the \(k^{\text{th}}\) eigenvalue of \(A\) and \(e_{n}(\rho_{1},\rho_{2},\dots,\rho_{d})\) is the \(n^{\text{th}}\) elementary symmetric polynomial which is defined as
\begin{gather}
	e_{n}(\rho_{1},\rho_{2},\dots,\rho_{d})\coloneqq\sum_{1\leqslant j_{1}<j_{2}<\cdots<j_{n}\leqslant d}\rho_{j_{1}}\rho_{j_{2}}\cdots\rho_{j_{n}}\,,\qquad n\leq d\,, \label{eqa2}
\end{gather}
with  \(e_{0}(\rho_{1},\rho_{2},\dots,\rho_{d})=1\) and  \(e_{n>d}(\rho_{1},\rho_{2},\dots,\rho_{d})=0\). While the expressions above provide a way to calculate a determinant of a \(d\times d\) matrix as a finite sum containing \(d\) terms, it is still a highly unpractical way to do so. In order to make life simpler consider the rank of the matrix \(A\) and the subsequent rank-nullity theorem, which in our case simply states that \(\rank(A-\rho \mathds{1}_{d})+\g_{A}(\rho)=d\), where \(\g_{A}(\rho)\) is the geometric multiplicity associated with the eigenvalue \(\rho\). Another useful fact is that \(\g_{A}(\rho)\leq\mu_{A}(\rho)\leq d\) where \(\mu_{A}(\rho)\) is the algebraic multiplicity of \(\rho\). Combining these two, we get that for \(\rho=0\) the following is true 
\begin{gather}
	d-\rank A\leq \mu_{A}(0)\,, \label{eqa3}
\end{gather}
which in simple words translates to: the matrix \(A\) has at least \(d-\rank A\) zero eigenvalues. Considering that 
\begin{gather}
	d=\sum_{i=1}^{k}\m_{A}(\rho_{i})=\m_{A}(0)+\m_{A}(\rho\neq0)\,, \qquad \m_{A}(\rho\neq0)\coloneqq\sum_{i=1}^{k}\m_{A}(\rho_{i})(1-\de_{\rho_{i},0})\,, \label{eqa4}
\end{gather}
where \(k\leq d\) is counting the distinct eigenvalues of \(A\), \eqref{eqa3} becomes
\begin{gather}
	\m_{A}(\rho\neq0)\leq\rank A\,,
\end{gather}
which means that, generally speaking, the maximum number of non-zero eigenvalues of the matrix \(A\) is equal to \(\rank A\). Now, we revisit \eqref{eqa2} and we notice that for any \(n\) such that \(\rank A<n\), we obtain \(e_{n}(\rho_{1},\rho_{2},\dots,\rho_{d})=0\). This is because we would have a product of \(n\) eigenvalues out of which only \(\rank A\) are non-zero in the best case for each term of the \(n\) sums over the eigenvalues. This allows us to rewrite \eqref{eqa1} in a much more ``convenient'' form
\begin{gather}
	\det(\rho \mathds{1}_{d}-A)=\sum_{n=0}^{\rank A}c_{n}\rho^{d-n}\,, \qquad c_{n}=(-1)^{n}e_{n}(\rho_{1},\rho_{2},\dots,\rho_{d})\,, \label{eqa6}
\end{gather}
where the series truncates significantly for small \(\rank A\). Finally, the Cayley-Hamilton theorem reads
\begin{gather}
	\sum_{n=0}^{\rank A}c_{n}A^{d-n}=0\,, \label{eqa7}
\end{gather}
which will be used extensively in this work.
\section{The ansatz method} \label{ansatz}
In this appendix, we will briefly review and marginally extend the ansatz method in \(d\) dimensions for comparison reasons. The standard flow equation at the action level reads
\begin{gather}
 	\pdv{S}{\la}=\mathcal{O}\,, \label{eq1}
\end{gather}
where \(\la\) is the deformation parameter and \(\mathcal{O}\) represents any deformation. A common starting ground includes Poincaré invariance, as well as minimally coupled to a non-singular background \(\tensor{g}{_\mu_\nu}\) theories. This leads to the well-known definition for the stress-tensor
\begin{gather}
	\tensor{T}{_\mu_\nu}\coloneqq\frac{-2}{\sqrt{-g}}\frac{\de S}{\de\tensor{g}{^\mu^\nu}}=-2\tensor{\Sigma}{_\mu_\nu}+\tensor{g}{_\mu_\nu}\mathcal{L}\,, \qquad \tensor{\Sigma}{_\mu_\nu}\coloneqq\pdv{\mathcal{L}}{\tensor{g}{^\mu^\nu}}\,. \label{eq2} 
\end{gather}
A natural assumption regarding the deformed theory is that for \(\la\rightarrow0\) one should retrieve some known seed Poincaré invariant density \(\mathcal{L}_{0}\).  A suitable general ansatz for the density \(\mathcal{L}\) is 
\begin{gather}
	\mathcal{L}=\mathcal{L}(x_{0},x_{1},\dots,x_{i},\dots,x_{D})\,, \label{eq7}
\end{gather}
where \(D\geqslant d\) is a positive integer. The index \(i\) that is running from \(0\) to \(D\) is a non-negative integer and is counting the number of inverse metrics that the corresponding (Lorentz scalar) element \(x_{i}\) contains. In other words
\begin{gather}
	x_{i}\coloneqq \text{all possible contractions of} \phantom{a} \tensor{X}{_{\mu_{1}}_{\nu_{1}}_{\cdots}_{\mu_{i}}_{\nu_{i}}} \phantom{a} \text{with} \phantom{a} \tensor{g}{^{\al_{1}}^{\be_{1}}}\cdots\tensor{g}{^{\al_{i}}^{\be_{i}}}\,, \label{eq8}
\end{gather} 
and the element \(x_{0}\) is understood to contain terms that do not couple to the background. The multi-indexed field \(\tensor{X}{_{\mu_{1}}_{\nu_{1}}_{\cdots}_{\mu_{i}}_{\nu_{i}}}\) corresponds to matter and is conventionally assumed to be independent of the metric. Based on the equation above, we will now categorise each case based on a finite number of building blocks and their corresponding index structure. For simplicity, we will always consider only one family of fields with a fixed number of indices i.e. we shall pick an element from the following set
\begin{gather}
	\qty{\tensor{X}{_{\mu_{1}}_{\mu_{2}}},\dots,\tensor{X}{_{\mu_{1}}_{\mu_{2}}_{\mu_{3}\cdots}}}\,, \label{8.1}
\end{gather}
and start building scalars with it. One reason for this choice is that any combination of fields with a different (or not) number of indices can be a special case of a field whose number of indices is equal to the sum of the number of indices of the independent fields that enter the combination we just mentioned. Another reason is that a potential combination of fields with a different number of indices, effectively increases the degrees of freedom of a theory which will result in significantly more complicated equations.  

Let us now consider only one family of fields that can be parametrised by \(\tensor{X}{_{\mu}_{\nu}}\), which is usually the case \cite{Bonelli:2018kik,Ferko:2022iru,Ferko:2022cix,Borsato:2022tmu}. Under this assumption the definition \eqref{eq8} simplifies to the following expression
\begin{gather}
	x_{i}= \tensor{g}{^{\mu_{\color{red}1\color{black}}}^{\nu_{\color{red}1\color{black}}}}\tensor{g}{^{\mu_{\color{blue}2\color{black}}}^{\nu_{\color{blue}2\color{black}}}}\color{Orange}\cdot\color{black}\cdot\cdot\tensor{g}{^{\mu_{\color{ForestGreen}i\color{black}}}^{\nu_{\color{ForestGreen}i\color{black}}}}\tensor{X}{_{\color{red}(\color{black} \mu_{\color{red}1\color{black}}}_{\color{blue}(\color{black}\nu_{\color{blue}2\color{black}}}}\tensor{X}{_{ \mu_{\color{blue}2\color{black}}\color{blue})\color{black}}_{\color{Orange}(\color{black}\nu_{\color{Orange}3\color{black}}}}\color{Orange}\cdot\color{black}\cdot\color{ForestGreen}\cdot\color{black}\tensor{X}{_{\mu_{\color{ForestGreen}i\color{black}}\color{ForestGreen})\color{black}}_{\nu_{\color{red}1\color{black}}\color{red})\color{black}}}\,. \label{eq9}
\end{gather}
Following a similar procedure as in \cite{Ferko:2022cix,ferko2021supersymmetry}, we now define the matrix \(X_{i}\) and \(x_{i}\) as
\begin{gather}
	\ten{\qty(X_{i})}{^\mu_\nu}\coloneqq\tensor{X}{^{\mu}_{\la_{1}}}\tensor{X}{^{ \la_{1}}_{\la_{2}}}\cdots\tensor{X}{^{\la_{i-1}}_{\nu}}\,, \qquad x_{i}\coloneqq\ten{\qty(X_{i})}{^\mu_\mu}=\tr(X_{i})\,,  \label{eq10}
\end{gather}
which leads to the following properties
\begin{gather}
	 \ten{\qty(X_{i+j})}{^\mu_\nu}=\ten{\qty(X_{i})}{^\mu_\la}\ten{\qty(X_{j})}{^\la_\nu}\,, \quad \ten{\qty(X_{i})}{^\mu_\nu}=\ten{\qty([X]^{i})}{^\mu_\nu}\,, \quad i,j\in\mathds{N}^{+}\,,  \label{eq11}
\end{gather}
and consequently we may rewrite \eqref{eq7} as
\begin{gather}
	\mathcal{L}=\mathcal{L}\qty(x_{0},\tr [X],\dots,\tr [X^{d}])\,, \label{eq12}
\end{gather}
where \(D\) is now reduced to \(d\) as a direct consequence of \eqref{eqa7}. This was first noticed in \cite{ferko2021supersymmetry} for the special antisymmetric case of \(\tensor{X}{_{\mu}_{\nu}}\) but for now, we choose to remain agnostic about any symmetries. Let us also note that these properties are generally true and independent of the nature of the deformation.
\subsection{Deformations with determinants}
In this section, we consider the following family of deformations
\begin{gather}
	\mathcal{O}=\int\dd[d]{x}\sqrt{-g}\hspace{0.1cm}\al[\det\tensor{T}{^\mu_\nu}]^{1/\be}\,, \label{eqcard}
\end{gather}
where \(\tensor{T}{^\mu_\nu}\) is the energy stress-tensor of the deformed theory \(\mathcal{L}\) and \(\al, \be\) are arbitrary constants. We should note that if \(\be>d\) the deformation is relevant, if \(\be=d\) it is marginal and finally, if \(\be<d\) the deformation is irrelevant. From \eqref{eq2}, we have
\begin{gather}
	\det\tensor{T}{^\mu_\nu}=\det(\mathcal{L}\tensor{\de}{^\mu_\nu}-2\tensor{\Sigma}{^\mu_\nu})\,. \label{eq3}
\end{gather}
The advantage of this ``one index upstairs, one index downstairs'' notation is that one does not have to contract with the (inverse) metric in order to calculate traces. This enables us to use formula \eqref{eqa6}
\begin{gather}
	\det\tensor{T}{^\mu_\nu}=\sum_{n=0}^{\rank\Sigma}c_{n}\mathcal{L}^{d-n}\,, \qquad c_{n}=(-1)^{n}e_{n}(\rho_{1},\rho_{2},\dots,\rho_{d})\,, \label{eq4}
\end{gather}
where it has become evident that the complexity of the expression is directly connected to the rank of \(\Sigma\). This was noticed for the first time in \cite{Bonelli:2018kik} for the \(\rank\Sigma=1\) case and explained in more detail in \cite{Ferko:2022iru}. Furthermore, since calculating the eigenvalues of \(\Sigma\) is not an easy task we can use the following formula 
\begin{gather}
	e_{n}(\rho_{1},\rho_{2},\dots,\rho_{d})=\frac{(-1)^{n}}{n!}B_{n}(-p_{1},-1!p_{2},-2!p_{3},\dots,-(n-1)!p_{n})\,, \qquad p_{n}\coloneqq\tr(\Sigma^{n})\,, \label{eq5}
\end{gather} 
where \(B_{n}\) is the \(n^{\text{th}}\) complete exponential Bell polynomial, thus \eqref{eq4} can take the following form
\begin{gather}
	\det\tensor{T}{^\mu_\nu}=\sum_{n=0}^{\rank\Sigma}c_{n}\qty[\tr(\Sigma),\tr(\Sigma^{2}),\dots,\tr(\Sigma^{n})]\mathcal{L}^{d-n}\,, \label{eq6}
\end{gather}
where we explicitly state the dependence of the coefficient \(c_{n}\) on the traces of the first \(n\) powers of \(\Sigma\). Looking back to \eqref{eq1} we observe that slowly, a non-linear yet first order partial differential equation is forming and the struggle to keep it under control comes not only from the rank of \(\Sigma\) but also from its power trace structure. Up to this point \(\mathcal{L}\) may contain an arbitrary number of matter fields from \eqref{8.1}. Proceeding with the two index family, a quick calculation shows
\begin{gather}
	\tensor{\Sigma}{^\mu_\nu}=\sum_{i=1}^{d}\frac{i}{2}\pdv{\mathcal{L}}{x_{i}}\tensor{(X_{i}+X^{\text{T}}_{i})}{^{\mu}_{\nu}}\,, \quad \tr(\Sigma)=\sum_{i=1}^{d}i\pdv{\mathcal{L}}{x_{i}}x_{i}\,, \quad \tensor{(X^{\text{T}}_{k})}{^{\mu}_{\nu}}\coloneqq\tensor{(X_{k})}{_{\nu}^{\mu}}\,, \label{detg1}
\end{gather}
and the trace of the \(n^{\text{th}}\) power of \(\Sigma\) is given by the following non-trivial expression
\begin{gather}
	\tr(\Sigma^{n})=\sum_{i_{1},\dots,i_{n}=1}^{d}\frac{i_{1}\cdots i_{n}}{2^{n}}\pdv{\mathcal{L}}{x_{i_{1}}}\cdots\pdv{\mathcal{L}}{x_{i_{n}}}\tr([X_{i_{1}}+X^{\text{T}}_{i_{1}}]\cdots[X_{i_{n}}+X^{\text{T}}_{i_{n}}])\,, \label{detg2}
\end{gather}
which reduces to 
\begin{gather}
	\tr(\Sigma^{n})=\hspace{-0.2cm}\sum_{i_{1},\dots,i_{n}=1}^{d}\qty[\prod_{k=1}^{n}i_{k}\pdv{\mathcal{L}}{x_{i_{k}}}]x_{i_{1}+\cdots+i_{n}}\,, \quad \tr(\Sigma^{n})=\hspace{-0.2cm}\sum_{i_{1},\dots,i_{n}=1}^{\lfloor \frac{d}{2} \rfloor}\qty[\prod_{k=1}^{n}2i_{k}\pdv{\mathcal{L}}{x_{2i_{k}}}]x_{2i_{1}+\cdots+2i_{n}}\,, \label{detg3}
\end{gather}
when \(\tensor{X}{_{\mu}_{\nu}}\) is symmetric and antisymmetric respectively. In the case of mixed symmetry, the expression is no longer a function of \(x_{i}\) and \(\partial_{x_{i}}\mathcal{L}\), hence this approach may not apply.
\subsection{Maxwell's theory}
We now follow with a minimal working example. Setting \(d=3\) and picking \(\tensor{X}{_{\mu}_{\nu}}\) to be antisymmetric leads to \(\mathcal{L}=\mathcal{L}(x_{0},x_{2})\) and it follows that \(\rank \Sigma=2\). Furthermore, we set \(\al=1\) and \(\be=2\) in \eqref{eqcard}. Then the combination of (\ref{eqa7}, \ref{eq1}, \ref{eqcard}) and \eqref{detg3} leads to
\begin{gather}
	\partial_{\la}\mathcal{L}=\sqrt{\mathcal{L}}\abs{\mathcal{L}-2x_{2}\partial_{x_{2}}\mathcal{L}}\,, \qquad \mathcal{L}_{0}=x_{2}\,, \label{pde1}
\end{gather}
where the absolute value comes from the square root. Picking either the positive \((+)\) or negative \((-)\) branch we reach the following solutions
\begin{gather}
	\mathcal{L}_{\text{d}}=\frac{2\qty(1\pm\sqrt{x_{2}}\la+\sqrt{1\pm2\sqrt{x_{2}}\la})}{\lambda^{2}}\,,\quad	\mathcal{L}_{\text{c}}=\frac{2\qty(1\pm\sqrt{x_{2}}\la-\sqrt{1\pm2\sqrt{x_{2}}\la})}{\lambda^{2}}\,. \label{pde2}
\end{gather}
The final step is to identify \(4 x_{2}\) with \(-\tensor{F}{_\mu_\nu}\tensor{F}{^\mu^\nu}\). It is worth stressing that only the \(\mathcal{L}_{\text{c}}\) set of solutions has a well-defined \(\la\rightarrow0\) limit whereas \(\mathcal{L}_{\text{d}}\) diverges to \(+\infty\) respectively. Furthermore, it is important to note that \eqref{pde2} is the deformed Lagrangian density of any matter action that is constructed via the antisymmetric \(\tensor{X}{_{\mu}_{\nu}}\) field with \(\mathcal{L}_{0}=x_{2}\). In other words, the identification \(\tensor{X}{_{\mu}_{\nu}}\leftrightarrow\tensor{F}{_\mu_\nu}\) is not necessary as far as the deformation is concerned.
\subsection{Quadratic and root deformations of sigma models}\label{sigan}
For this section we set \(d=2\) and we consider the deformation given by \eqref{opp2d} i.e.
\begin{gather}
	\mathcal{O}=\int\dd[2]{x}\sqrt{-g}\hspace{0.1cm}2b_{0}\qty{-\det\tensor{T}{^\mu_\nu}+\frac{b_{2}}{\la'}\tensor{T}{^\mu_\mu}-\frac{b_{2}(1+b_{2})}{{\la'}^{2}}}\,, \label{ttbar}
\end{gather}
where \( \la'\coloneqq2\la b_{0}\). Notice that for \(b_{0}=1/2\) and \(b_{2}=0\) the operator reduces to \(T\overline{T}\). Another interesting albeit marginal operator is \cite{Ferko:2022cix}
\begin{gather}
	\mathcal{R}=\int\dd[2]{x}\sqrt{-g}\hspace{0.05cm}\dfrac{b_{\text{r}}}{\sqrt{2}}\sqrt{\tensor{T}{^\mu_\nu}\tensor{T}{^\nu_\mu}-\frac{1}{2}(\tensor{T}{^\mu_\mu})^{2}}\,, \label{ttlike} 
\end{gather}
which is usually referred as the root-\(T\overline{T}\) deformation. As in the previous case, \(b_{\text{r}}\) is arbitrary and dimensionless. We will now proceed with the deformation of sigma models considering the two operators above. The non-linear \(\s\)-model is given by the following density
\begin{gather}
	\mathcal{L}_{0}=-\frac{1}{2}\tensor{g}{^\mu^\nu}\tensor{\Phi}{_\mu_\nu}-V+\dfrac{\mathcal{T}}{\sqrt{-g}}\,, \qquad \tensor{\Phi}{_\mu_\nu}\coloneqq\tensor{A}{_i_j}\partial_{(\mu}\phi^{i}\partial_{\nu)}\phi^{j}\,, \label{incond}
\end{gather}
where \(V\) is an arbitrary potential and \(\mathcal{T}\) corresponds to any topological term. From there we can identify \(\tensor{\Phi}{_\mu_\nu}\leftrightarrow\tensor{X}{_\mu_\nu}\) where \(\tensor{X}{_\mu_\nu}\) is now symmetric and consequently \(x_{1}\leftrightarrow\tensor{g}{^\mu^\nu}\tensor{\Phi}{_\mu_\nu}\). This leads to the following ansatz
\begin{gather}
	\mathcal{L}=\mathcal{L}(\Omega,x_{0},x_{1},x_{2})\,, \qquad \Omega\coloneqq\sqrt{-g}\,. \label{santz}
\end{gather}
Then, the combination of (\ref{eqa7}, \ref{eq1}, \ref{detg3}) and \eqref{ttbar} leads to the following, highly non-trivial partial differential equation
\begin{gather}
	\begin{aligned}
	\partial_{\la'}\mathcal{L}=(2x_{2}-x_{1}^{2})(\partial_{x_{1}}\mathcal{L}+2x_{1}\partial_{x_{2}}\mathcal{L})^{2}-\qty(\mathfrak{d}\mathcal{L}-\dfrac{b_{2}}{\lambda' })^{2}-\frac{b_{2}}{{\la'}^{2}}\,,   \label{ttbarpde}
		\end{aligned}
\end{gather}
 where \(\mathfrak{d}\coloneqq1-x_{1}\partial_{x_{1}}-2x_{2}\partial_{x_{2}}+\Omega\partial_{\Omega}\). Similarly, for \eqref{ttlike} we get
\begin{gather}
	\partial_{\m}\mathcal{L}=b_{\text{r}}\sqrt{2x_{2}-x_{1}^{2}}(\partial_{x_{1}}\mathcal{L}+2x_{1}\partial_{x_{2}}\mathcal{L})\,, \label{ttlikepde}
\end{gather}
where \(b_{\text{r}}\) has absorbed a factor of \(\pm\) coming from the square root and \(\m\) is the dimensionless deformation parameter for the root-\(T\overline{T}\) operator. An interesting fact about the equation above is that all \(\sqrt{-g}\) related terms cancel exactly, something that is clearly not true in \eqref{ttbarpde}. Equation \eqref{ttlikepde} can be solved exactly \cite{Ferko:2022cix} i.e. 
\begin{gather}
	\mathcal{L}=\mathcall{L}_{0}\qty(\Omega,x_{0},x_{1,\m},x_{2,\m})\,, \label{ttlikesol1}
\end{gather}
where
\begin{gather}
\begin{aligned}
	&x_{1,\m}=x_{1}\cosh(b_{\text{r}}\m)+\sqrt{2x_{2}-x_{1}^{2}}\sinh(b_{\text{r}}\m)\,,\\&x_{2,\m}=x_{2}\cosh(2b_{\text{r}}\m)+x_{1}\sqrt{2x_{2}-x_{1}^{2}}\sinh(2b_{\text{r}}\m)\,,
	\end{aligned} \label{ttlikesol2}
\end{gather}
with \(x_{2,\m}-x_{1,\m}^{2}=x_{2}-x_{1}^{2}\) and \(\mathcall{L}_{0}\) being any function. On the other hand, equation \eqref{ttbarpde} together with \eqref{incond} gives
\begin{gather}
	\mathcal{L}_{\pm}=\dfrac{1+b_{2}}{\la'}-\dfrac{1}{2\widetilde{\la'}}\qty(1\pm\sqrt{1+2\widetilde{\la'}x_{1,\m}+2{\widetilde{\la'}}^{2}\qty(x_{1}^{2}-x_{2})})+\dfrac{\mathcal{T}}{\sqrt{-g}}\,, \label{solsigma}
\end{gather}
where \(\widetilde{\la'}\coloneqq\la'(1-V\la')\) and \(x_{1}\) is promoted to \(x_{1,\m}\) in accordance with \eqref{ttlikesol1} and \eqref{ttlikesol2}. One can easily check that \eqref{solsigma} satisfies \eqref{ttbarpde} and \eqref{ttlikepde} simultaneously. Moreover, the limit \(\m\rightarrow0\) is always regular, but the limit \(\la\rightarrow0\) diverges to \(\pm\infty\) with the exception of \(b_{2}=0\) and \(\mathcal{L}_{+}\), or \(b_{2}=-1\) and \(\mathcal{L}_{-}\). In the first case we recover \eqref{incond} with \(x_{1}\rightarrow x_{1,\m}\) and in the second \(x_{1,\m}/2+\mathcal{T}/\sqrt{-g}\). Finally, the two flows commute since \eqref{ttbarpde} retains its form after the change of coordinates: \(x_{1}\rightarrow x_{1,\m}\) and \(x_{2}\rightarrow x_{2,\m}\).
\section{Deformed energy spectrum} \label{derspec}
In this final appendix, we will derive the energy spectrum of the operator \eqref{2ddef} taking into account only the first constraint \eqref{con1}. The operator now becomes 
\begin{gather}
	\mathcal{O}=\int\dd[2]{x}\sqrt{-g}\hspace{0.05cm}\qty{-2b_{0}\det\tensor{T}{^\mu_\nu}+\frac{b_{2}}{\la}\tensor{T}{^\mu_\mu}+\frac{b_{3}}{\la^{2}}}\,. \label{spectt}
\end{gather}
As in \cite{Zamolodchikov:2004ce}, we place our theory on a Euclidean cylinder of circumference \(R\) with coordinates \((\ta,x)\sim(\ta,x+R)\) and \(\ta=it\). After Wick rotating, \eqref{eq1} can be written as
\begin{gather}
	\partial_{\la}\log\mathcal{Z}_{\text{E}}=-\expval{\mathcal{O}_{\text{E}}}\,, \label{spec1}
\end{gather}
where the subscript E denotes the Euclidean version of the corresponding quantity. Please do keep in mind that \eqref{spec1} is generally true, and is independent of the operator \eqref{spectt} or the cylinder. In our case it follows
\begin{gather}
	\partial_{\lambda}E_{n}(\la,R)=2b_{0}R\ev{\det\tensor*{T}{_{\text{E}}^\mu_\nu}}{n}-\dfrac{b_{2}R}{\la}\ev{\tensor*{T}{_{\text{E}}^\mu_\mu}}{n}-\dfrac{b_{3}R}{\la^{2}}\,,
\end{gather}
and since\footnote{For a detailed derivation of the pressure term one may refer to \cite{Kruthoff:2020hsi}.}
\begin{gather}
	\ev{\tensor{T}{_\ta_\ta}}{n}=-\dfrac{E_{n}(\la,R)}{R}\,,\quad 	\ev{\tensor{T}{_\ta_x}}{n}=-\dfrac{iP_{n}(\la,R)}{R}\,, \quad 	\ev{\tensor{T}{_x_x}}{n}=-\pdv{E_{n}(\la,R)}{R}\,,
\end{gather}
we reach the following equation
\begin{gather}
\partial_{\lambda}E_{n}=2b_{0}\qty(E_{n}\pdv{E_{n}}{R}+\dfrac{P_{n}^{2}}{R})+\dfrac{b_{2}}{\la}\qty(E_{n}+R\pdv{E_{n}}{R})-\dfrac{b_{3}R}{\la^{2}}\,, \label{flowspec}
\end{gather}
where the \(\la, R\) dependence is suppressed. It is rather interesting that \eqref{flowspec} does not seem to have the form of the inviscid Burgers' equation. Nevertheless, it can still be solved via the following dimensionless quantities \cite{Giveon:2017nie, Jiang:2019epa}
\begin{gather}
	A_{n}(u)\coloneqq RE_{n}(\la,R)\,, \qquad \mathcal{P}_{n}\coloneqq RP_{n}(\la,R)\,,  \qquad u\coloneqq\dfrac{\la}{R^{2}}\,,
\end{gather}
where \(\mathcal{P}_{n}\) is a constant. This choice reduces \eqref{flowspec} to an ordinary differential equation
\begin{gather}
(1+2b_{2}+4b_{0}uA_{n})A_{n}'-2b_{0}(\mathcal{P}_{n}^{2}-A_{n}^{2})+\dfrac{b_{3}}{u^{2}}=0\,, \qquad A_{n}'\coloneqq\dv{A_{n}}{u}\,. \label{flowode}
\end{gather}
The solutions read
\begin{gather}
	E^{\pm}_{n}(\la',R)=-\dfrac{R\qty(1+2b_{2})}{2\la'}\pm\sqrt{\qty(\dfrac{\mathcal{P}_{n}}{R})^{2}+\qty(\dfrac{R\qty(1+2b_{2})}{2\la'})^{2}+\dfrac{2b_{0}\qty(b_{3}R^{2}+\la'\mathcal{E}_{n})}{{\la'}^{2}}}\,, \label{solsprec}
\end{gather}
where \(\mathcal{E}_{n}\) is determined via an appropriate initial condition and \(\mathcal{P}_{n}=2\pi p_{n}\), with \(p_{n}\in\mathds{Z}\) on the cylinder.
\section{Some deformations and their coefficients} \label{bees}
In this appendix, we will provide the deforming operators in five, six and seven dimensions along with a table that includes their higher order coefficients. In short, the deformation in five dimensions is given by
	\begin{gather}
	\begin{aligned}
		\small\mathcal{O}=\int\dd[5]{x}f\hspace{0.05cm}\Bigg{\{}\la^{3}b_{0}\tr T^{5}+\la^{3}b_{1}\tr T^{4}\tr T+\la^{3} b_{2}\tr T^{3}\tr T^{2}+\la^{3} b_{3}\tr T^{3}(\tr T)^{2}+&\\[-0.2cm]+\la^{3} b_{4}(\tr T^{2})^{2}\tr T+\la^{3} b_{5}\tr T^{2}(\tr T)^{3}+\la^{3} b_{6}(\tr T)^{5}+\la^{2} b_{7}\tr T^{4}+&\\[0.2cm]+\la^{2} b_{8}\tr T^{3}\tr T+\la^{2} b_{9}(\tr T^{2})^{2}+\la^{2} b_{10}\tr T^{2}(\tr T)^{2}+&\\[0.2cm] +\la^{2} b_{11}(\tr T)^{4}+\la b_{12}\tr T^{3}+ \la b_{13}\tr T^{2}\tr T +\la b_{14}(\tr T)^{3}+&\\+\beta\qty(\tr T^{2}-\dfrac{1}{4}(\tr T)^{2})+\dfrac{\beta_{l}}{\la}\tr T +\dfrac{\beta_{c}}{\la^{2}}&\Bigg{\}}\,, \label{5dopdef}
	\end{aligned}
\end{gather}
in six dimensions by
\begin{gather}
	\begin{aligned}
		\small\mathcal{O}=\int\dd[6]{x}f\hspace{0.05cm}\Bigg{\{}\la^{4}b_{0}\tr T^{6}+\la^{4}b_{1}\tr T^{5}\tr T+\la^{4} b_{2}\tr T^{4}\tr T^{2}+\la^{4} b_{3}\tr T^{4}(\tr T)^{2}+&\\[-0.2cm]+\la^{4} b_{4}(\tr T^{3})^{2}+\la^{4} b_{5}\tr T^{3}\tr T^{2}\tr T +\la^{4} b_{6}\tr T^{3}(\tr T)^{3}+\la^{4} b_{7}(\tr T^{2})^{3}+&\\[0.2cm]+\la^{4} b_{8}(\tr T^{2})^{2}(\tr T)^{2}+\la^{4} b_{9}\tr T^{2}(\tr T)^{4}+\la^{4} b_{10}(\tr T)^{6}+\la^{3} b_{11}\tr T^{5}+&\\[0.2cm]+\la^{3} b_{12}\tr T^{4}\tr T+\la^{3} b_{13}\tr T^{3}\tr T^{2}+\la^{3} b_{14}\tr T^{3}(\tr T)^{2}+ \la^{3} b_{15}(\tr T^{2})^{2}\tr T +&\\[0.2cm] +\la^{3} b_{16}\tr T^{2}(\tr T)^{3}+\la^{3} b_{17}(\tr T)^{5}+\la^{2} b_{18}\tr T^{4}+\la^{2}b_{19}\tr T^{3}\tr T+&\\[0.2cm]+\la^{2}b_{20}(\tr T^{2})^{2}+\la^{2} b_{21}\tr T^{2}(\tr T)^{2}+\la^{2} b_{22}(\tr T)^{4}+\la b_{23}\tr T^{3}+&\\+ \la b_{24}\tr T^{2}\tr T +\la b_{25}(\tr T)^{3}+\beta\qty(\tr T^{2}-\dfrac{1}{5}(\tr T)^{2})+\dfrac{\beta_{l}}{\la}\tr T +\dfrac{\beta_{c}}{\la^{2}}\Bigg{\}}&\,, \label{6dopdef2}
	\end{aligned}
\end{gather}
and finally in seven dimensions the deformation reads
\begin{gather}
	\begin{aligned}
		\small\mathcal{O}=\int\dd[7]{x}f\hspace{0.05cm}\Bigg{\{}\la^{5}b_{0}\tr T^{7}+\la^{5}b_{1}\tr T^{6}\tr T+\la^{5} b_{2}\tr T^{5}\tr T^{2}+\la^{5} b_{3}\tr T^{5}(\tr T)^{2}+&\\[-0.2cm]+\la^{5} b_{4}\tr T^{4}\tr T^{3}+\la^{5} b_{5}\tr T^{4}\tr T^{2}\tr T +\la^{5} b_{6}\tr T^{4}(\tr T)^{3}+\la^{5} b_{7}(\tr T^{3})^{2}\tr T+&\\[0.2cm]+\la^{5}b_{8}\tr T^{3}(\tr T^{2})^{2}+\la^{5}b_{9}\tr T^{3}\tr T^{2}(\tr T)^{2}+\la^{5} b_{10}\tr T^{3}(\tr T)^{4}+&\\[0.2cm]+\la^{5} b_{11}(\tr T^{2})^{3}\tr T+\la^{5} b_{12}(\tr T^{2})^{2}(\tr T)^{3}+\la^{5} b_{13}\tr T^{2}(\tr T)^{5}+\la^{5} b_{14}(\tr T)^{7}+&\\[0.2cm]+\la^{4} b_{15}\tr T^{6}+\la^{4} b_{16}\tr T^{5}\tr T+\la^{4} b_{17}\tr T^{4}\tr T^{2}+\la^{4} b_{18}\tr T^{4}(\tr T)^{2}+&\\[0.2cm]+\la^{4} b_{19}(\tr T^{3})^{2}+\la^{4} b_{20}\tr T^{3}\tr T^{2}\tr T+\la^{4} b_{21}\tr T^{3}(\tr T)^{3} +\la^{4} b_{22}(\tr T^{2})^{3}+&\\[0.2cm]+ \la^{4} b_{23}(\tr T^{2})^{2}(\tr T)^{2} +\la^{4} b_{24}\tr T^{2}(\tr T)^{4}+\la^{4} b_{25}(\tr T)^{6}+\la^{3} b_{26}\tr T^{5}+&\\[0.2cm]+\la^{3} b_{27}\tr T^{4}\tr T+\la^{3} b_{28}\tr T^{3}\tr T^{2}+ \la^{3} b_{29}\tr T^{3}(\tr T)^{2} +\la^{3} b_{30}(\tr T^{2})^{2}\tr T+&\\[0.2cm] +\la^{3} b_{31}\tr T^{2}(\tr T)^{3}+\la^{3} b_{32}(\tr T)^{5}+\la^{2}b_{33}\tr T^{4}+\la^{2}b_{34}\tr T^{3}\tr T+&\\[0.2cm]+\la^{2} b_{35}(\tr T^{2})^{2}+\la^{2} b_{36}\tr T^{2}(\tr T)^{2}+\la^{2} b_{37}(\tr T)^{4}+ \la b_{38}\tr T^{3}+&\\+\la b_{39}\tr T^{2}\tr T+\la b_{40}(\tr T)^{3}+\beta\qty(\tr T^{2}-\dfrac{1}{6}(\tr T)^{2})+\dfrac{\beta_{l}}{\la}\tr T +\dfrac{\beta_{c}}{\la^{2}}\Bigg{\}}&\,. \label{7dopdef}
	\end{aligned}
\end{gather}
The coefficients \(b_{i}\) are given in the table \ref{table:b1's} below, whereas \(\be_{l}\) and \(\be_{c}\) are given by \eqref{linconbffe}.
\begin{table}[H]
	\centering
	{\renewcommand{\arraystretch}{1.2}%
		\begin{tabular}{c|c|c|c}
			\(b_{i}\) &	\(d=5\)  & \(d=6\) & \(d=7\) \\	
			\hline
			\(b_{1}\) & \(-\frac{15}{16}a^{3}\) & \(-\frac{24}{25}a^{4}\) & \(-\frac{35}{36}a^{5}\) \\
			\(b_{2}\)& \(-\frac{5}{6}a^{3}\) & \(-\frac{3}{4}a^{4}\) & \(-\frac{7}{10}a^{5}\)  \\
			\(b_{3}\) & \(\frac{25}{48}a^{3}\) & \(\frac{51}{100}a^{4}\) & \(\frac{91}{180}a^{5}\)\\
			\(b_{4}\) & \(\frac{15}{32}a^{3}\) & \(-\frac{1}{3}a^{4}\) & \(-\frac{7}{12}a^{5}\)\\
			\(b_{5}\) & \(-\frac{85}{384}a^{3}\) & \(\frac{4}{5}a^{4}\) & \(\frac{35}{48}a^{5}\)\\
			\(b_{6}\) & \(\frac{53}{3072}a^{3}\) & \(-\frac{74}{375}a^{4}\) & \(-\frac{161}{864}a^{5}\)\\
			\(b_{7}\) & \(\frac{5}{4}a^{2}z\) & \(\frac{1}{8}a^{4}\) & \(\frac{35}{108}a^{5}\)\\
			\(b_{8}\) & \(-\frac{5}{6}a^{2}z\)& \(-\frac{51}{200}a^{4}\)  & \(\frac{7}{24}a^{5}\)\\
			\(b_{9}\) & \(-\frac{5}{8}a^{2}z\) & \(\frac{329}{5000}a^{4}\) & \(-\frac{91}{216}a^{5}\)\\
			\(b_{10}\) & \(\frac{15}{32}a^{2}z\) & \(-\frac{1349}{375000}a^{4}\) & \(\frac{217}{3888}a^{5}\)\\
			\(b_{11}\) & \(-\frac{35}{768}a^{2}z\) & \(\frac{6}{5}a^{3}z\) & \(-\frac{35}{288}a^{5}\)\\
			\(b_{12}\) & \(\frac{5}{3}az^{2}\) & \(-\frac{9}{10}a^{3}z\) & \(\frac{161}{1728}a^{5}\)\\
			\(b_{13}\) & \(-\frac{5}{8}az^{2}\) & \(-a^{3}z\) & \(-\frac{1183}{77760}a^{5}\)\\
			\(b_{14}\) & \(\frac{5}{96}az^{2}\) & \(\frac{11}{25}a^{3}z\) & \(\frac{859}{1399680}a^{5}\)\\ 	
		\end{tabular}\qquad
		\begin{tabular}{c|c|c}
			\(b_{i}\) & \(d=6\) & \(d=7\) \\	
			\hline	
			\(b_{15}\) & \(\frac{9}{20}a^{3}z\) & \(\frac{7}{6}a^{4}z\) \\
			\(b_{16}\) & \(-\frac{41}{250}a^{3}z\) & \(-\frac{14}{15}a^{4}z\) \\
			\(b_{17}\) & \(\frac{131}{12500}a^{3}z\) & \(-\frac{7}{8}a^{4}z\) \\
			\(b_{18}\) & \(\frac{3}{2}a^{2}z^{2}\) & \(\frac{7}{16}a^{4}z\) \\
			\(b_{19}\) & \(-\frac{4}{5}a^{2}z^{2}\) & \(-\frac{7}{18}a^{4}z\) \\
			\(b_{20}\) & \(-\frac{3}{4}a^{2}z^{2}\) & \(\frac{7}{9}a^{4}z\) \\
			\(b_{21}\) & \(\frac{21}{50}a^{2}z^{2}\) & \(-\frac{49}{324}a^{4}z\) \\
			\(b_{22}\) & \(-\frac{17}{500}a^{2}z^{2}\) & \(\frac{7}{48}a^{4}z\) \\
			\(b_{23}\) & \(2az^{3}\) & \(-\frac{7}{32}a^{4}z\) \\
			\(b_{24}\) & \(-\frac{3}{5}az^{3}\) & \(\frac{119}{2592}a^{4}z\) \\
			\(b_{25}\) & \(\frac{1}{25}az^{3}\) & \(-\frac{497}{233280}a^{4}z\) \\
		\end{tabular}\qquad
		\begin{tabular}{c|c}
			\(b_{i}\) & \(d=7\) \\	
			\hline	
			\(b_{26}\) & \(\frac{7}{5}a^{3}z^{2}\) \\
			\(b_{27}\) & \(-\frac{7}{8}a^{3}z^{2}\) \\
			\(b_{28}\) & \(-\frac{7}{6}a^{3}z^{2}\) \\
			\(b_{29}\) & \(\frac{7}{18}a^{3}z^{2}\) \\
			\(b_{30}\) & \(\frac{7}{16}a^{3}z^{2}\) \\
			\(b_{31}\) & \(-\frac{7}{54}a^{3}z^{2}\) \\
			\(b_{32}\) & \(\frac{91}{12960}a^{3}z^{2}\) \\
		\end{tabular}\qquad
		\begin{tabular}{c|c}
			\(b_{i}\) & \(d=7\) \\	
			\hline	
			\(b_{33}\) & \(\frac{7}{4}a^{2}z^{3}\) \\
			\(b_{34}\) & \(-\frac{7}{9}a^{2}z^{3}\) \\
			\(b_{35}\) & \(-\frac{7}{8}a^{2}z^{3}\) \\
			\(b_{36}\) & \(\frac{7}{18}a^{2}z^{3}\) \\
			\(b_{37}\) & \(-\frac{35}{1296}a^{2}z^{3}\) \\
			\(b_{38}\) & \(\frac{7}{3}az^{4}\) \\
			\(b_{39}\) & \(-\frac{7}{12}az^{4}\) \\
			\(b_{40}\) & \(\frac{7}{216}az^{4}\) \\
	\end{tabular}}
	\centering
	\caption{All higher order coefficients are listed. We have defined \(b_{0}\eqqcolon a^{d-2}\) and \(\beta\eqqcolon d z^{d-2}/2\) for notational transparency.}
	\label{table:b1's}
\end{table}
\noindent Let us briefly provide some insight about the table above. First, one notices that \(b_{1}\) is equal to \(-d(d-2)/(d-1)^{2}a^{d-2}\), and this gives the correct answer in all the cases that we have studied. Moreover, the number of variables \(b_{i}\) is given by 
\begin{gather}
\sum_{i=0}^{d}p(i)-5\,,
\end{gather}
where \(p(i)\) is equal to the number of integer partitions of \(i\). Finally, their evaluation in \(d\) dimensions is accompanied by a discouraging fact. When the \(d\)-dimensional deformation is considered, the leading order term, which is proportional to \(\tr T^{d}\), introduces the ``highest order'' variable \(y_{d^{2}}\), in other words any \(y_{i}\) with \(i>d^{2}\) does not enter the analysis. Due to the Cayley-Hamilton theorem \eqref{eqa7}, the independent variables are \(y_{i}\) with \(1\leq i\leq d\) which means that in our case one needs to re-express the \(d(d-1)\) variables in terms of the independent ones. This approach is quite tedious and is not very productive for large \(d\). Surprisingly, the solution \eqref{solldbb} eliminates this problem completely. To fully understand this fact, let us once more consider the source of the term \(y_{d^{2}}\) but now from the massive gravity point of view. Taking a closer look to the solution, we observe that the term proportional to \(\det Y_{1}\) contains all variables from \(y_{1}\) up to and including \(y_{d}\). In the process of evaluating the stress-tensor, we vary this term with respect to the background vielbein and the contribution from \(y_{d}\) is something proportional to \(Y_{d}\). Then, evaluating \(\tr T^{d}\) leads to the term \(y_{d^{2}}\). It is now obvious that the extra \(d(d-1)\) variables are coming entirely from \(\det Y_{1}\). But we mentioned multiple times that this term decouples completely from the background and becomes a \(\la\)-dependent part of the seed theory, therefore by pushing \(\det Y_{1}\) out of the massive gravity action and evaluating the stress-tensor, the \(d(d-1)\) variables identically vanish from the \(d\)-dimensional deforming operator. Their effect is essentially reduced to the addition of the term \(-db_{0}\mathfrak{c_{1}}^{\!\!\!d}\det Y_{1}/\la^{2}\) on the left-hand side of equation \eqref{pde}.
\renewcommand{\refname}{\centering{References}}  
\addcontentsline{toc}{section}{\protect\numberline{}\protect\hspace*{-\cftsecnumwidth}References}
\bibliographystyle{bibstyle}
\bibliography{sample}
\end{document}